\documentclass[useAMS,usenatbib]{mn2e}
\usepackage{amsmath}
\usepackage{txfonts}
\usepackage{url}
\usepackage{natbib}
\usepackage{subfigure}
\usepackage{times}
\usepackage{epstopdf}
\usepackage{afterpage}
\usepackage{graphicx}
\usepackage{amsmath, amssymb}

\numberwithin{equation}{section}
\newcommand{\ha}{$\text{H}\alpha$ }
\newcommand{\vi}{\text{\texttt{v}} }
\newcommand{\gaia}{\textit{Gaia}}
\title[GS-TEC: the \textit{Gaia} Spectrophotometry Transient Events Classifier]{GS-TEC: the \textit{Gaia} Spectrophotometry Transient Events Classifier}
\author[Nadejda Blagorodnova et al.]
{Nadejda Blagorodnova$^1$\thanks{E-mail:nblago@ast.cam.ac.uk},
Sergey E. Koposov $^{1,2}$ \thanks{E-mail:koposov@ast.cam.ac.uk},
\L ukasz Wyrzykowski $^{1,3}$,  
\newauthor Mike Irwin $^1$ and Nicholas A. Walton $^1$ 
\\$^{1}$Institute of Astronomy (IoA), University of Cambridge, Madingley Road, Cambridge, CB3 0HA United Kingdom\\
$^2${Moscow MV Lomonosov State University, Sternberg Astronomical Institute, Moscow, 119992, Russia}\\
$^3$Warsaw University Observatory, Al. Ujazdowskie 4, 00-478 Warszawa, Poland}

\begin{document}

\date{Accepted 2014 April 25.  Received 2014 April 24; in original form 2014 January 8}

\pagerange{\pageref{firstpage}--\pageref{lastpage}} \pubyear{2014}

\maketitle

\label{firstpage}

\begin{abstract}
We present an algorithm for classifying the nearby transient objects detected by the \gaia{} satellite. The algorithm will use the low-resolution spectra from the blue and red spectro-photometers on board of the satellite. Taking a Bayesian approach we model the spectra using the newly constructed reference spectral library and literature-driven priors. We find that for magnitudes brighter than 19 in \gaia{} $G$ magnitude, around 75\% 
of the transients will be robustly classified. The efficiency of the algorithm for SNe type I is higher than 80\% for magnitudes $G\leq$18, dropping to approximately 60\% at magnitude $G$=19. For SNe type II, the efficiency varies from 75 to 60\% for $G\leq$18, falling to 50\% at $G$=19. 
The purity of our classifier is around 95\% for SNe type I for all magnitudes. For SNe type II it is over 90\% for objects with $G \leq$19. 
GS-TEC also estimates the redshifts with errors of $\sigma_z \le$ 0.01 and epochs with uncertainties $\sigma_t \simeq$ 13 and 32 days for type SNe I and SNe II respectively. GS-TEC has been designed to be used on partially calibrated \gaia{} data. However, the concept could be extended to other kinds of low resolution spectra classification for ongoing surveys.
\end{abstract}


\section{Introduction}

The study of transient phenomena is a field of increasing interest: for example, the observations of type Ia Supernovae (SNe) have lead to the discovery 
of the accelerated expansion of the Universe (\cite{1999ApJ...517..565P}, \cite{Riess1998}) and have played a fundamental role in the discovery of Dark Energy. 
Furthermore the investigation of transient phenomena at multiple wavelengths have lead to a better understanding of SNe progenitors \citep{2009ARA&A..47...63S} and modelling of the explosion mechanisms.

The era of large transient surveys has just begun with, for example, the Palomar Transient Factory (PTF, \cite{Rau2009}), Pan-STARRS \citep{2002SPIE.4836..154K}, and  Catalina Realtime Transient Survey (CRTS, \cite{CRTS2011}). \gaia{}, the ESA cornerstone mission \citep{GaiaPerryman}, whilst primarily an astrometry mission, will have a significant ability in revealing the transient universe. \gaia{}, will provide highly accurate parallaxes for over a billion stars. In addition, it will provide a wealth of additional information about each star: full six dimensional astrometric parameters; and astrophysical parameters such as effective temperature, surface gravity, metallicties and reddening. Since \gaia{} will observe each point of the sky around 70 times on average, it will, over the nominal mission length of 5 years, detect many thousands of new transient events. Indeed \gaia{} is expected to discover between 6000 and 7000 new SNe (\cite{Belokurov2003}, \cite{Altavilla2012}), thus several SNe each day, down to a limiting magnitude of \gaia{} $G$=20 which for SNe events corresponds to a redshift limit $z \lesssim 0.14$. 

The \gaia{} photometric science alerts system \citep{Wyrzykowski2012} will perform the detection, classification and dissemination of the alerts on transient events to the scientific community. The alerts system will process all data from \gaia{}, on a daily basis, as soon as the data is downloaded. In the simplest case, it will issue alerts based on flux changing by more than a defined magnitude threshold. GS-TEC is a standalone module using the \gaia{} photometric and spectrophotometric data allows the alerts system assign a classification type and a classification probability to each alert. This module is one of the three different classification modules that the photometric science alerts intends to use for classification purposes. The spectroscopic classification result provided by GS-TEC will be published along with photometric data (lightcurve of the event as detected by Gaia) and environment of the transient based on catalogue search. However, it will be the only one providing information on SNe subtypes and their parameters. The description of the full photometric science alerts pipeline and first results is aimed to be released in a separate paper.

The alert stream will be non-proprietary and will be distributed via public on-line services. The time to release the alerts is still to be determined, but it will take between 24 and 48 hours since the on-board observation. Its main goal is to provide information to enable targeted selection for follow-up and to filter the objects according to their scientific relevance. At this point it becomes essential to provide a reliable classification algorithm that can provide information on the nature of the event, e.g. AGN, variable star, SNe (plus its type), in addition to providing parameters, such as redshift, or epoch to maximum brightness for the case of slowly evolving objects like SNe. Other type of events such as Cataclysmic Variables or Tidal Disruption Flares are also relevant for Gaia classification scheme, however, the (almost) lack of broad features in their spectra makes them a difficult target for a low resolution spectral classification only. Therefore, in the present context they will be considered as part of the black body-like population, which is included in the classification.

The importance of having a real-time automated detection and classification framework has been already pointed out by the teams of PTF: \cite{Brink2013} and \cite{Bloom2012}, and Catalina \citep{Djorgovski2011} synoptic surveys. An average night may receive several hundreds of potential alerts, which need to be processed in nearly real time in order to characterize them and select the most interesting targets for rapid follow-up.

This paper describes the classification algorithm developed to enable the prototyping of SNe events from \gaia{}, where the primary information source is the \gaia{} low resolution spectrophotometric data. The paper has the following outline: Section~\ref{sec:gaia-bprp} summarizes the most relevant characteristics of \gaia{} Blue Photometer (BP) and Red Photometer (RP), Section~\ref{sec:library} describes the assembly of the reference spectral template library, Section~\ref{sec:algorithm} summarizes the method employed. Sections~\ref{sec:results} and~\ref{sec:pessto} contains the results of applying the classifier on ground-based observations of transient objects. The discussion of the results is contained in Section~\ref{sec:discussion} whilst summary and conclusions are presented in Section~\ref{sec:conclusions}.

\section{\gaia{} spectrophotometry}\label{sec:gaia-bprp}

\begin{figure}
\includegraphics[width=1.1\columnwidth]{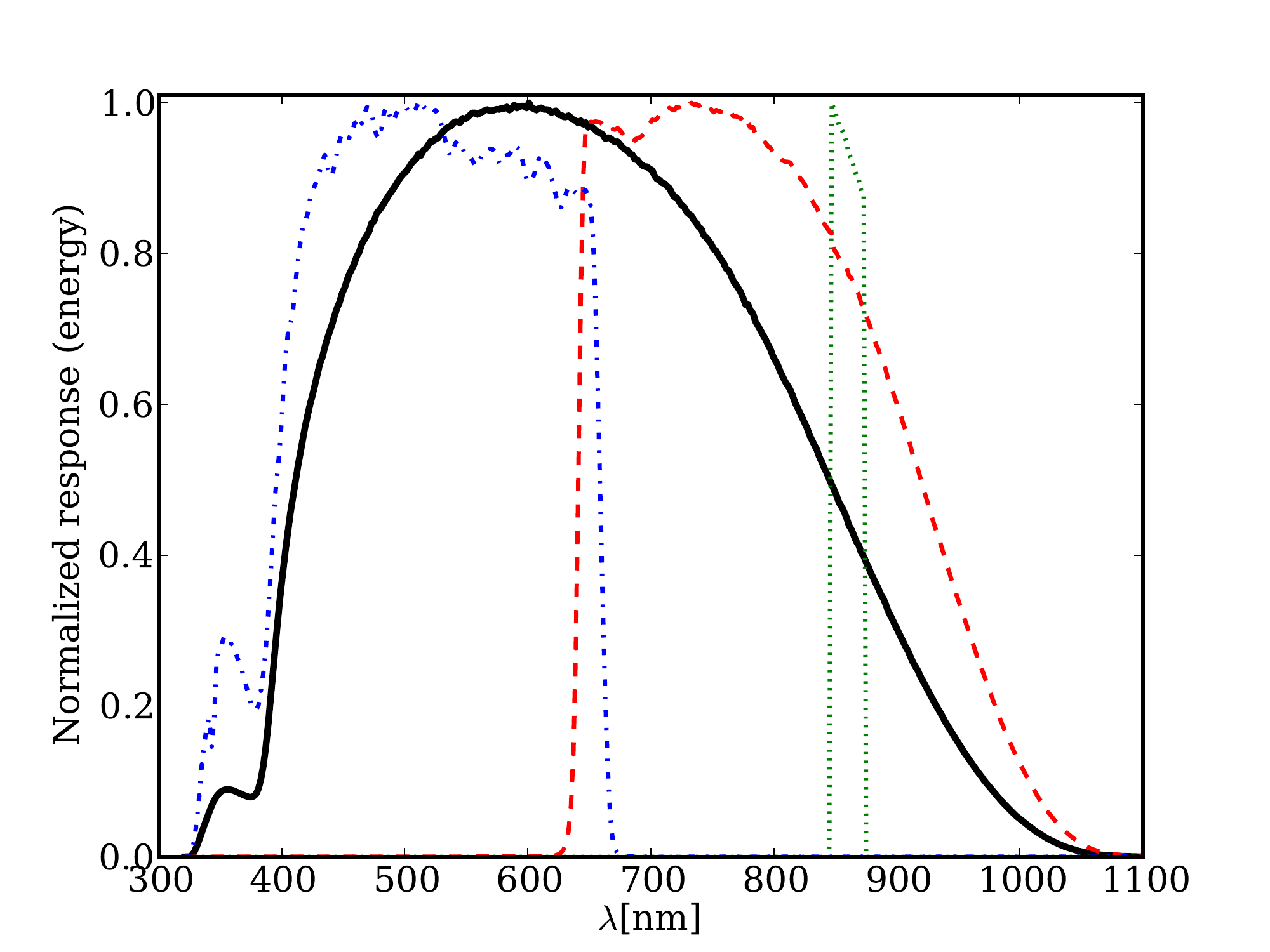}
\caption{\gaia{} magnitude pass-band (solid black line), pass-bands of the blue (dot-dashed blue line) and red spectrophotometers (dot dashed red line). The normalized throughput of the \gaia{} RVS instrument is shown by the dotted green line.}
\label{fig:bprp_disp_curve}
\end{figure}

\gaia{} has four different passbands: $G; G_{BP}; G_{RP}$; and  $G_{RVS}$.  Their wavelength coverage is shown in figure \ref{fig:bprp_disp_curve}. Each passband response is a convolution of the optical response curves and the quantum efficiency curves for each CCD type. The prisms that disperse the light for the two photometric bands have coatings that work as low-pass and high-pass filters for the BP and RP \citep{Jordi2010}.

The \gaia{} $G$ magnitude corresponds to the unfiltered light from the astrometric field, which covers almost all the optical range (330$-$1050\,nm). Its accuracy decreases from 0.3 millimag at G=12 to 20 millimag at G=20 \citep{2012ApSS.tmp...68D}, which makes it possible to monitor the brightness variability history for virtually all the objects observed by \gaia{}.  Alerts on transient events will be raised when new objects or statistically significant changes in magnitude are detected.

\begin{figure*}
\centering
\includegraphics[width=0.99\textwidth]{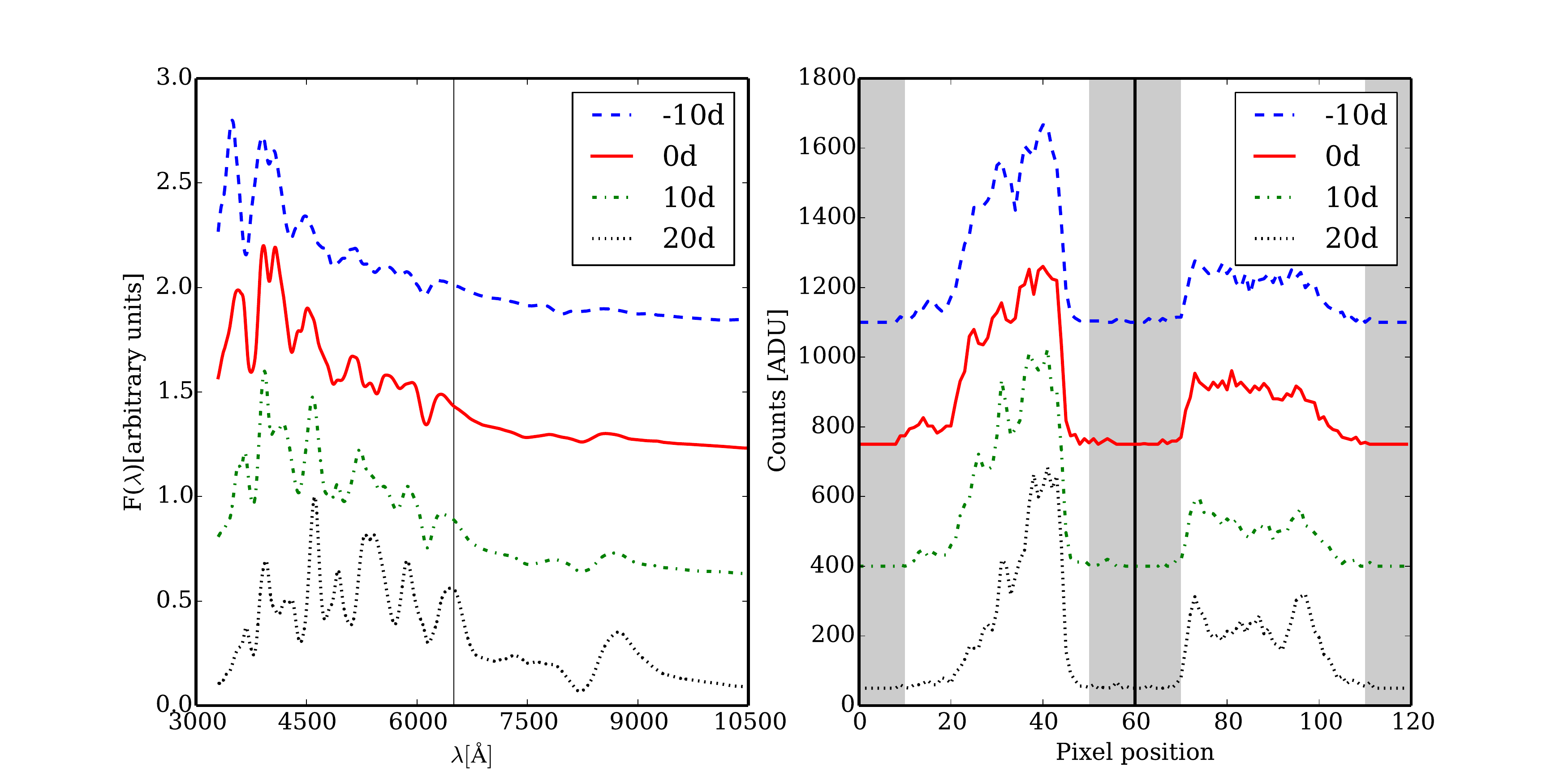}
  \caption{{\it Left}: Medium resolution spectra (10\,$\AA$ pixel$^{-1}$) of template spectra for SNe Ia at -10, 0, +10 and +20 days relative to maximum brightness in the visual band. The thin line at 6500\,$\AA$ is a visual guide to distinguish between the area covered by the red and by the blue spectrophotometers. {\it Right}: same spectra converted into a BP/RP high signal-to-noise low-resolution format into counts /ADUs). Right hand side: 60 pixels from BP (330$-$680nm), left hand side: 60 pixels from RP (640$-$1050nm). The grey areas are discarded in our analysis as they hardly carry any information. }
  \label{fig:sn_bprp}
\end{figure*}

The BP and RP cover the optical ranges 330$-$680\,nm and 640$-$1050\,nm respectively and provide low resolution spectrophotometry with sampling ranging from 4 to 32\,nm\,pixel$^{-1}$  for BP and 7 to 15\,nm\,pixel$^{-1}$ for RP. According to the target apparent magnitude, 2-dimensional or 1-dimensional windows ( a 2-dimensional window binned in the across-scan direction) will be allocated around the point-like sources in the CCD, resulting in two- or one-dimensional sets of measurements per object. The integrated calibrated fluxes for each band are transformed to $G_{BP}$ and $G_{RP}$ magnitudes while the individual fluxes per pixel are transformed to a common instrument pixel reference frame. Flux and wavelength calibrations are then applied to obtain properly calibrated spectra. Figure \ref{fig:sn_bprp} shows a comparison between ground-based low resolution SNe spectra and \gaia{}-like BP and RP equivalents. 

The \gaia{} radial velocity spectrometer (RVS) covers the $G_{RVS}$ wavelength range 847$-$874\,nm, to observe part of the spectra around the Ca~II triplet lines. This part of the spectrum is dispersed by a grating providing a resolution power of $\sim$11500 for stars brighter than $G\sim 17$ magnitude. This relatively shallow limit precludes using the data from this instrument in our transient alert analysis.

\section{Transient spectra training set }\label{sec:library}

One of the many ways to approach a classification and parametrization problem is to rely on models or templates,
which can be used as a training set.
These objects provide an important reference point to compare the incoming data against. This section explains how the library was created from a set of spectral sources.

\subsection{Sources of spectral training set}

The reference libraries were constructed by collating transient spectra from several sources of observed and model (template) spectra.

The observed spectra mainly come from the PTF \citep{Rau2009} and are available via WiseRep \footnote{\url{http://www.weizmann.ac.il/astrophysics/wiserep}} resource \citep{Yaron2012}. The other two sources are spectra from the CfA Supernova Data Archive\footnote{\url{http://www.cfa.harvard.edu/supernova/SNarchive.html}} for SNe type Ia spectra and the Asiago SNe catalogue \citep{Asiago1999}.

The template spectra for galaxies and AGN were taken from the SWIRE library, \citep{Polletta2007}. The HILIB Stellar Set of 131 stellar spectra with types from O5 to M2I \citep{StellarLibrary} were used as stellar templates. SNe templates are based on data from Peter Nugent \footnote{\url{http://supernova.lbl.gov/~nugent/nugent_templates.html}} and E. Hsiao SNe Ia templates \citep{2007ApJ...663.1187H}. Finally we included in our template library a set of black body spectra with temperatures ranging from 3000\,K to 30000\,K to emulate objects with (almost) featureless spectra.

\subsection{Standardization process of observed spectra} \label{sec:standard}

When building a library using ground-based observed spectra we are faced with the problem of a set of heterogeneous, non-standardised data. For  use as a reference set for classification the data needs to be as homogeneous as possible. 

The key steps of our homogenisation process are: correction for redshift to bring all spectra to rest-frame wavelengths; correction for reddening effects; and an edge correction to extend the irregular wavelength coverage of observed spectra to the fixed wavelength range 330 to 1050\,nm covered by Gaia. 
This standardisation procedure implies that we have to assign to every library spectrum a set of  parameters which characterize the observation 
   \begin{align}
    \theta = (t, z, A_{V}), \label{eq:theta}
    \end{align}
where $t$ is the epoch of the observation measured in days before/after the maximum brightness in a particular photometric band, $z$ is the redshift and $ A_{V}$ is the extinction in magnitudes in the $V$ passband. In a perfect world the spectra of objects of the same spectral type and with the same parameters $\theta$  would be alike. However, in practice objects of the same type will exhibit differences depending on factors such as metallicity, luminosity, mass outflow, the density of the surrounding medium and so on, which will modify both the general spectral shape and the individual strength of absorption and emission lines. Since it is impractical to model all possible effects, we include them in our analysis using a statistical approach.

Of the three parameters included in $\theta$, two, namely the redshift and the approximate epoch of explosion (for SNe) are usually either provided in the spectra repositories, or can be found in published literature. However, the extinction values are usually unknown and the effect of the extinction on the spectral shape can be significant.

To estimate the extinction knowing the redshift of the object and epoch of the observation, we compare the observed spectra with reddened templates at the same epoch. A range of values of $A_V$ are applied to the template spectrum in 0.1dex steps using the Cardelli extinction law \citep{1989ApJ...345..245C} and $R_V=3.1$. Template and observed spectra are then compared using a $\chi^2$ statistic to find the $A_V$ value providing the best match.

\begin{figure}
 \centering
 \hspace{-1cm}
 \includegraphics[width=1.1\columnwidth]{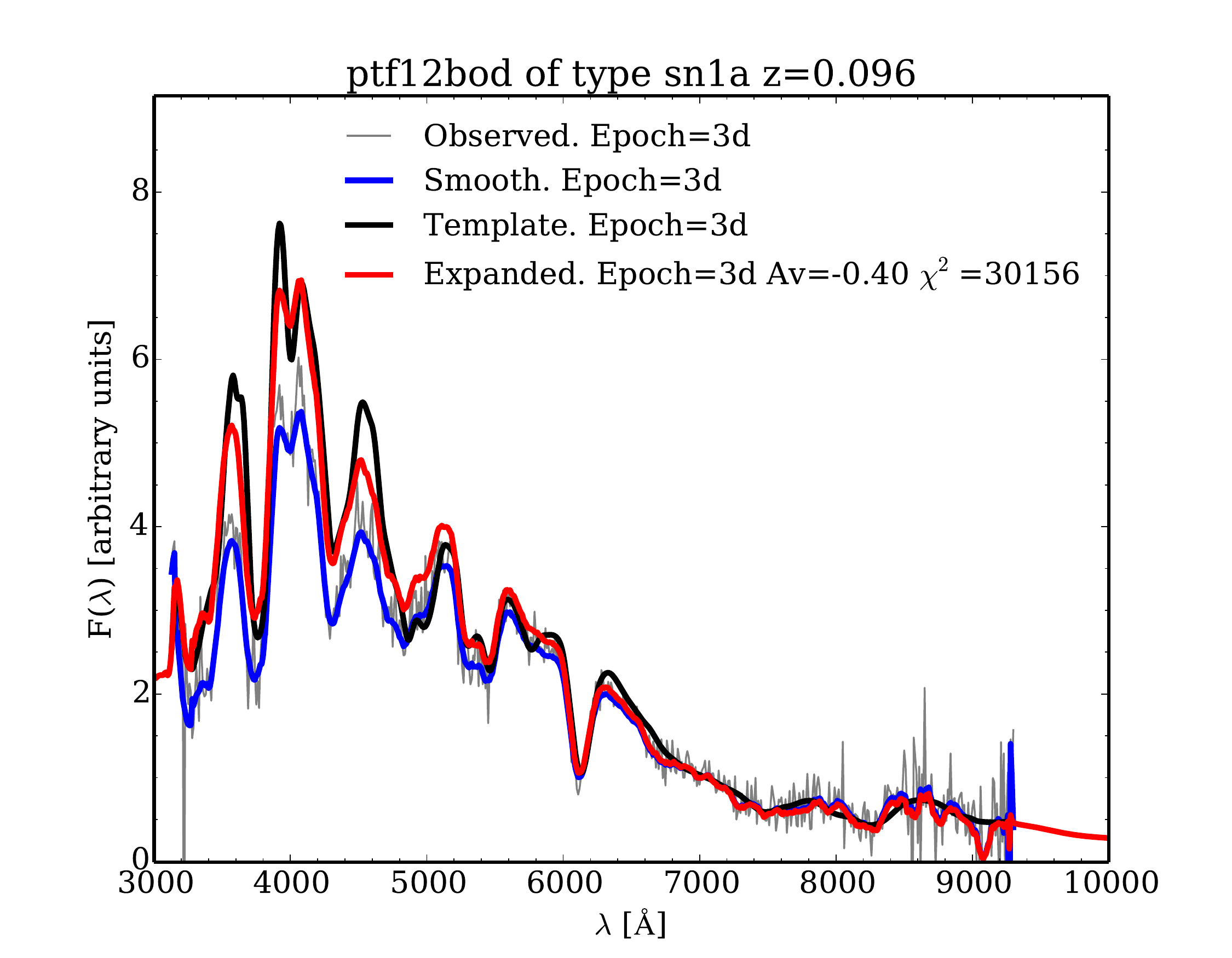}
 \caption{Illustration of the extinction-correction process aiming at dereddening spectra by comparison with templates. The figure shows the individual steps of this procedure: observed rest-frame spectra (grey thin line); observed spectra smoothed by a Gaussian filter with $\sigma=$ 100 $\AA$(blue line); template of the same spectral type and epoch (black line); smoothed observed spectra, extinction-corrected and extended around edges spectrum (red line). The extinction correction applied was $A_V$=0.4\,mag as determined from the $\chi^2$ fit with the template.}\label{fig:standardization}
 \end{figure}

After de-reddening, the observed spectra are extrapolated, if necessary, using the best template match to cover all the \gaia{} wavelength range. Sharp transitions between observed and template spectra are minimised by using 20\,nm overlaps to ensure smoothness. This procedure generally has very little impact on the simulated Gaia spectra as the instrument response is much lower in the blue and red ends. Figure \ref{fig:standardization} shows an illustration of the extinction determination and correction procedure. 
The result of the standardization process is a set of spectra of transients, extinction-corrected, rest-frame corrected and fully defined over the wavelength range 300$-$1100\,nm.

 \begin{figure}
 \centering
  \hspace{-1cm}
 \includegraphics[width=1.1\columnwidth]{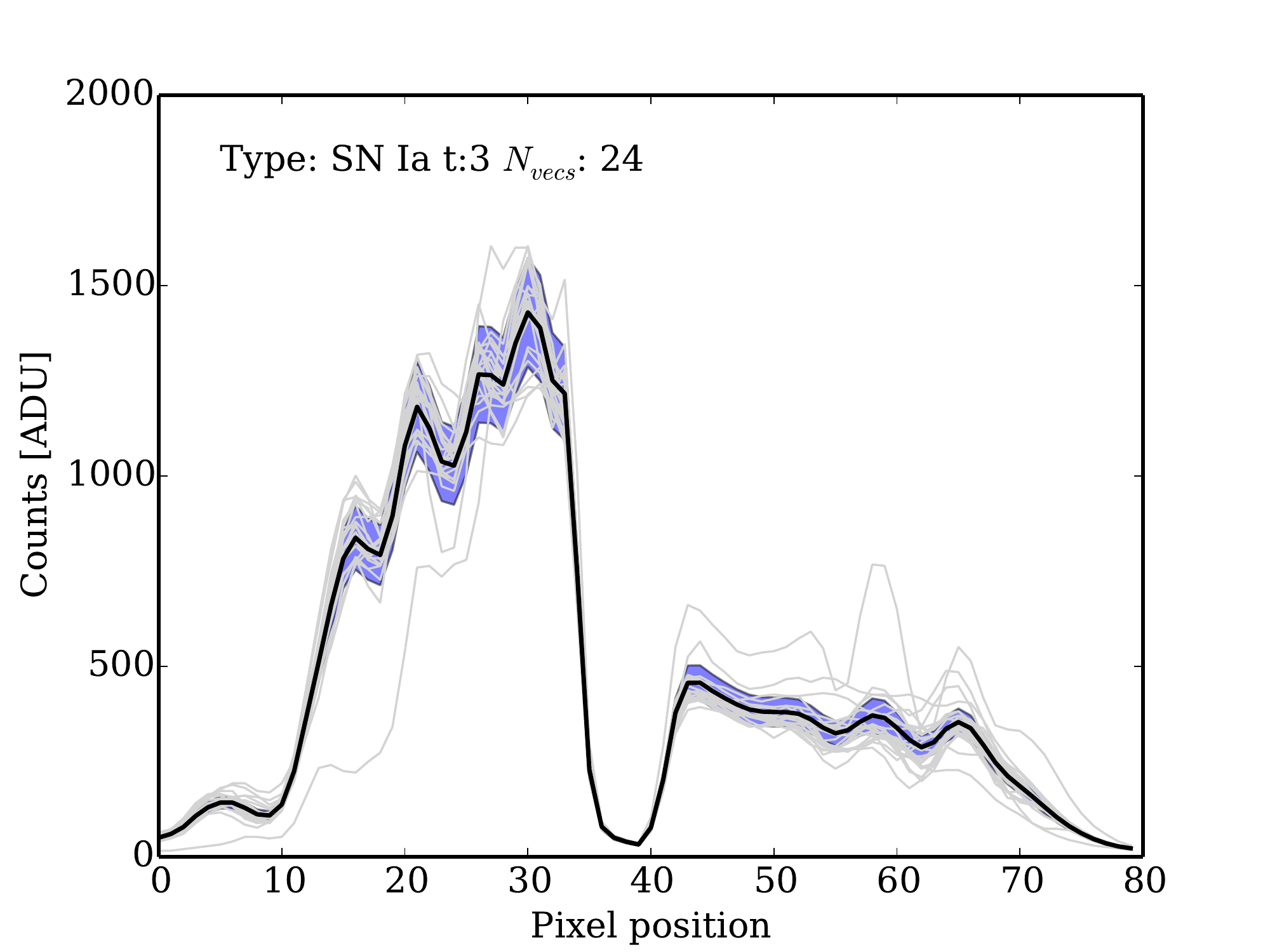}
 \caption{Representation of the medium BP/RP spectrum of a SNe type Ia at rest-frame as seen 3 days after maximum brightness with no extinciton. The black line shows the median spectrum of 24 input vectors. The blue region shows the 1$\sigma$ region, while the grey lines show individual SNe Ia spectra, showing considerable variance in the details of the spectra. The amplitude of this variation is accounted for in the model intrinsic dispersion.}\label{fig:Variability}
 \end{figure}

\subsection{Library parametrization} \label{red:parametrization}
The spectral reference library can be interpreted as a forward model: given a particular object type and $\theta$ parameter we can predict the spectral shape of the object in the \gaia{} BP/RP space. The synthesis of the predicted spectra requires selecting the standardized spectra with the right type and epoch, redshifting and reddening it according to $\theta$ and \textit{downgrading} to \gaia{} format. This last transformation requires an intermediate step, which computes the number of photons per unit wavelength, per unit time, and per unit surface area, photons $\text{s}^{-1} \text{m}^{-2} \text{nm}^{-1}$ from the original spectra $f_{\lambda}$ in units of $\text{erg } \text{cm}^{-2} \AA^{-1} \text{s}^{-1}$. The transformation is given by 
\begin{equation}
\text{N}_p(\lambda) = E/E_p = \frac{ \text{F}(\lambda) \times 1 \times 10^{-12} \times \lambda}{hc}
\label{eq:photon_transf}
\end{equation}
where $F(\lambda)$ is the flux in $\text{erg }\text{cm}^{-2} \AA^{-1} \text{s}^{-1}$, $h$ is Planck's constant in Js and $c$ is the speed of light in m s$^{-1}$.

Transformation to a Gaia-like format is done by an internal \gaia{} DPAC (Data Processing and Analysis Consortium) simulation module designed to create BP/RP spectra called \textit{XpSim} (Brown, A., private communication). This module convolves the spectra with the optical response and BP and RP QE curves (figure \ref{fig:bprp_disp_curve}) to generate low-resolution spectra as would be provided by \gaia{}.

As noted previously, despite our homogenisation procedure there will still be some 
intrinsic variation among spectra with the same types and epochs and we will account for that in a statistical way. To obtain a measure of the intrinsic variance of spectra for objects with same type and parameters we group those with the same parameters $\theta$ and then compute the median spectrum and its variance. Figure \ref{fig:Variability} shows a median BP/RP spectrum of a SNe type Ia 3 days after maximum. For this particular case 24 standardized spectra were used to compute the median spectrum. The standard deviation of the spectra in each pixel represents the model intrinsic dispersion at that pixel.

Finally, as the observed spectra that we include in our library do not include all possible epochs, we fill the gaps in the epoch dimension by linearly interpolating the spectra and their variances to a grid of epochs with 1 day spacing.

\subsection{Transient numbers estimation}
The classification process requires some prior information on the expected number of transients for each class. In order to estimate the number of alerts for a given object type in the \gaia{} survey for a limiting magnitude $G$=19, we have to take several assumptions. From \cite{Altavilla2012} we estimate an optimistic number of 7000 SNe. \cite{Li2011} provides the internal rates for a magnitude-limited survey with a 30 day cadence. The non-SNe rates are much less certain, and the given numbers are expected to be updated during the mission.

The expected number of AGN is given by \cite{Mignard2012}. From \cite{MacLeod2012} we estimate that only a fraction of 0.001 of AGN will vary more than 0.5 mag in a period of 30 days. If we consider this variability as a threshold to trigger an alert, the expected number of detected AGN is around 500 objects. 
The number of black body objects (BB), focuses mainly on very young core collapse SNe (which are a minority given the Gaia cadence), Tidal Disruption Flares and Novae. Tidal Disruptions are rather rare events, so we will focus on the case of Classical and Dwarf Novae. Their number is inferred from the fraction relative to SNe in the predictions realized for the PTF survey \citep{Rau2009}. In order to account for the difference in cadence among the two surveys, which is 5 days for PTF and 30 days for Gaia, we assumed a uniform distributions for their time to decay and computed the difference in the detected fraction.

Variable star contamination is difficult to predict, as no specific rates have been computed so far for \gaia{}, according to their variability type, amplitudes and periods. Their discovery along the mission will populate the Gaia internal reference catalogues. The most suitable to be mistaken as transients are the variables with long period and high amplitude, such as Miras. According to \cite{Eyer2000}, in total \gaia{} will observe around 140 000 Mira variables. These objects will be potential contaminants at the initial stages of the mission, specially during the first 6 months. We could assume therefore that around 1400 objects will be observed at magnitude 19 during this time. A summary of these numbers is displayed in table \ref{tab:rates}. However, it worth noting that the detection rates for each type of objects are going to evolve with time, along with the transient discovery history of \gaia{}. After several months, Gaia-specific rates will replace the \textit{a priori} estimates.

\section{Classification algorithm}
\label{sec:algorithm}

\subsection{Data description}
The data to be classified is in the form of BP and RP one-dimensional vectors, each 60 elements long. We assume that the data vector has been calibrated for the effects of light dispersion due to different positions in the CCD as this process is included in the data reduction pipeline. In some parts of the spectrum the quantum efficiency is very low (grey area in in Figure \ref{fig:sn_bprp}) and we have ignored them in the classification process leaving 40 central pixels for each instrument. For convenience these two vectors are concatenated to a single data vector $\{d_i\}$ 80 elements long. Each element of the vector $d_i$ contains the photon counts for that pixel and we also have the estimate of the measurement error $e_i$ which together constitute the data $D= \{d_i,e_i\}$. The model for these data is determined by the type of the object $M$ and the realised model for the spectral vector {$f(i| \theta, M)$} plus the intrinsic variance attached to the model spectra {$\omega_i$}. Two other input parameters are available: \vi a visibility flag which is set to 1 if the object is bright enough to be detected by \gaia{} and 0 otherwise; and $m_G$ the apparent magnitude of the objects at the time of the \gaia{} observation. A summary of the notation used is given in Table~\ref{tab:notation}.

\begin{table}
\begin{tabular}{p{1.8cm} p{5.7cm}}

\hline\hline
$d=\{d_1..d_N\}$ & data vector, where $d_i$ is the number of photon counts in position $i$ \\
$e=\{e_1..e_N\}$ & error vector, where $e_i$ is the error on the observed value $d_i$  \\
$D=\{d, e\}$ & input data vector with uncertainties which represents the spectra to be classified \\
$M$ & model, represents the type (or class) of the event: \{SN Ia, SN Ibc, SN IIP, SN IIn, AGN, STAR, BB\} \\
$\theta=(z,t,A_V)$ & parameters for model M \\
$f(i|\theta,M)$ & function which predicts the flux in position $i$ given the parameters and the model M \\ 
$\omega_i$ & uncertainty on the model $f(i|\theta,M)$ due to intrinsic model dispersion \\
$G_{M}$ & absolute magnitude for objects of type M. \\
$\sigma_M$ & standard deviation on the absolute magnitude $G_0$. \\
$m_G$ & \textit{Gaia} apparent magnitude of the input vector. \\
$m_T$ & theoretically predicted apparent magnitude for transient with model M with parameters $\theta$ \\
$\sigma_T$ & theoretically predicted sigma for magnitude $m_T$ \\
$m_{lim}$ & limiting magnitude for the \gaia{} survey.\\ 
\vi & condition that the object is detectable by \gaia{} \\ 
\hline
\end{tabular}
\caption{Summary of the notation used.}
\label{tab:notation}
\end{table}

\subsection{Bayesian classification method} \label{sec:model}

The classification is essentially a model selection problem and we have used an adaptation of a time-series model selection method \citep{Bailer-Jones2012}. The goal is to compute the probability of each model provided: the observed data, the measurement error and the prior information on the frequency of different models (object classes) $P(M)$, displayed in table\ref{tab:rates}. As output we expect an array of normalized posterior probabilities for each model (object class). The probability for each individual model $M$ is be given by

\begin{equation}
P(M|D, m_G, \vi) = \frac{P(D, m_G, \vi|M) P(M)} { \int{ P(D,m_G,\vi|M)P(M)} d M} \text{ .}
\label{eq:model_prob}
\end{equation}

 \begin{figure*}
\centering
\mbox{
\subfigure{\includegraphics[width=0.9\columnwidth]{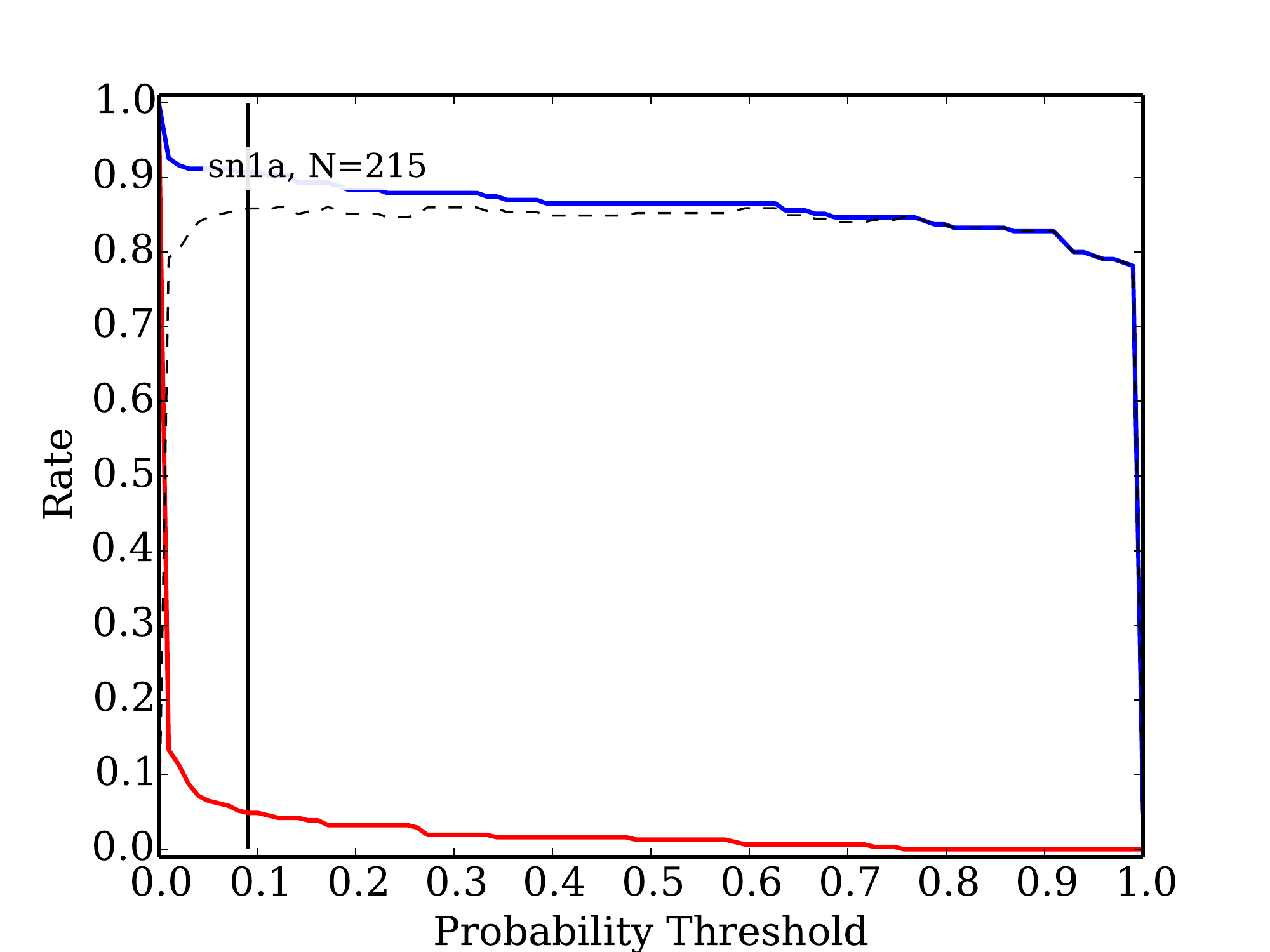} 
\quad
\subfigure{\includegraphics[width=0.9\columnwidth]{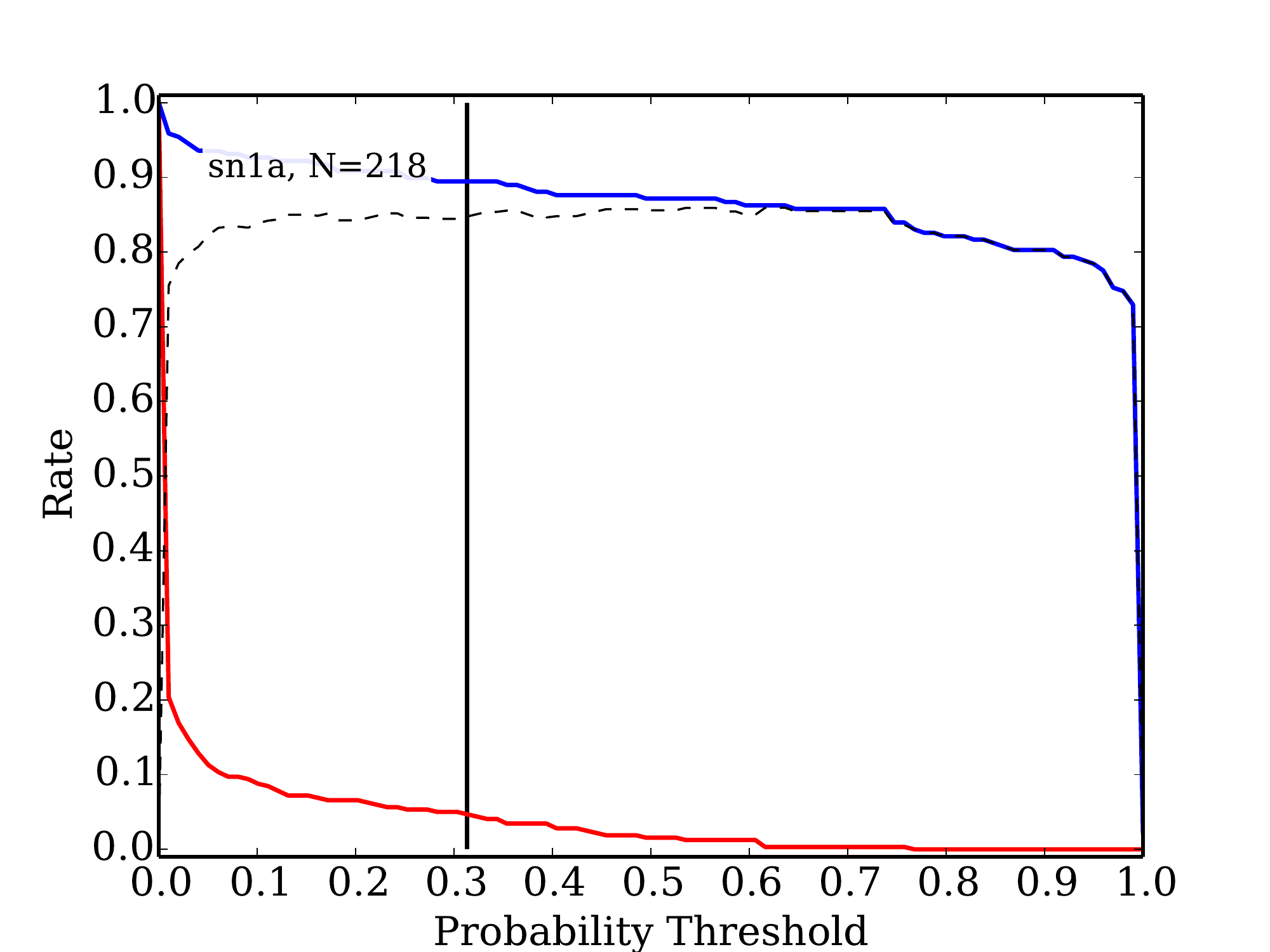} }} }
\mbox{
\subfigure{\includegraphics[width=0.9\columnwidth]{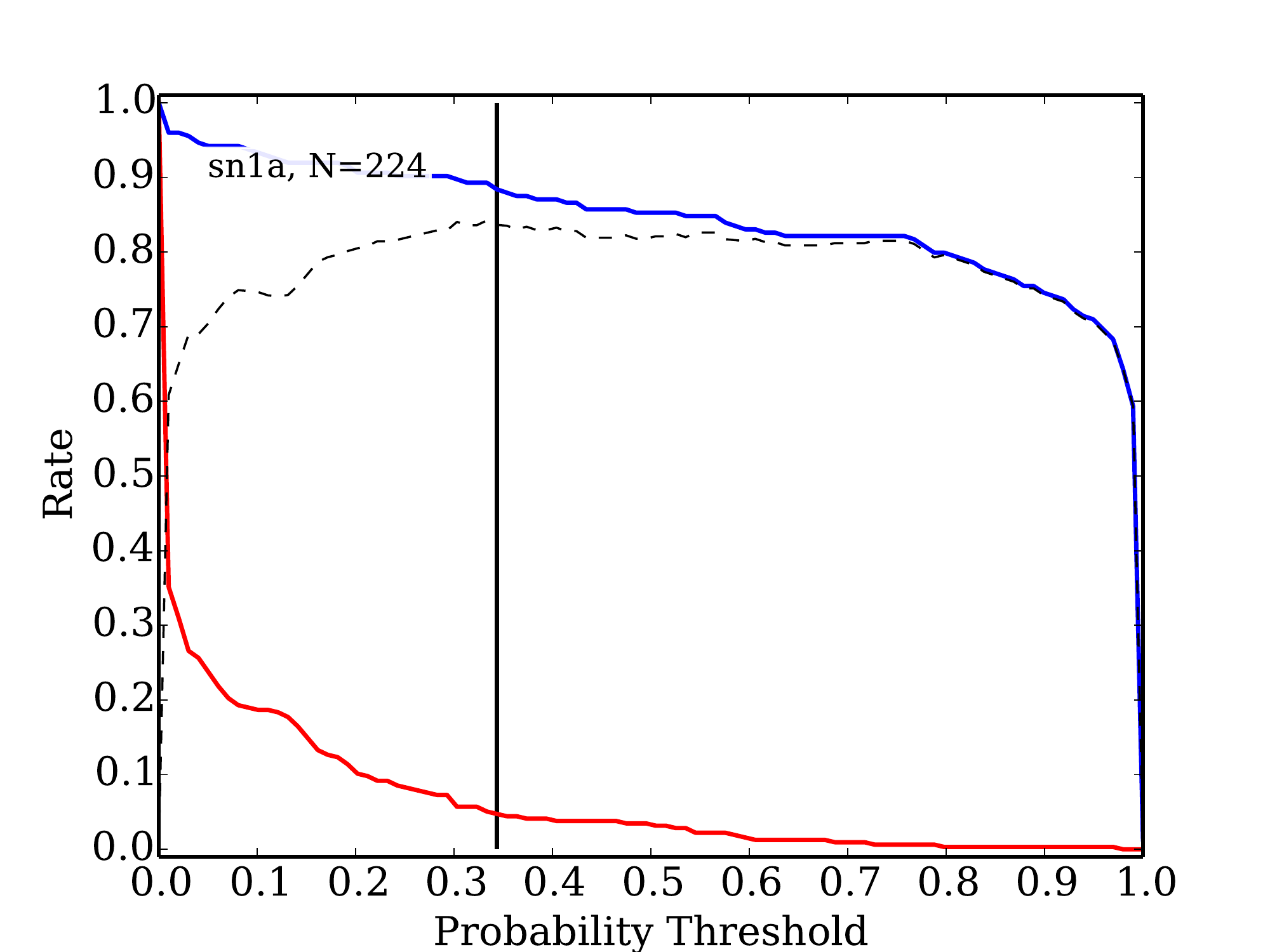}
\quad
\subfigure{\includegraphics[width=0.9\columnwidth]{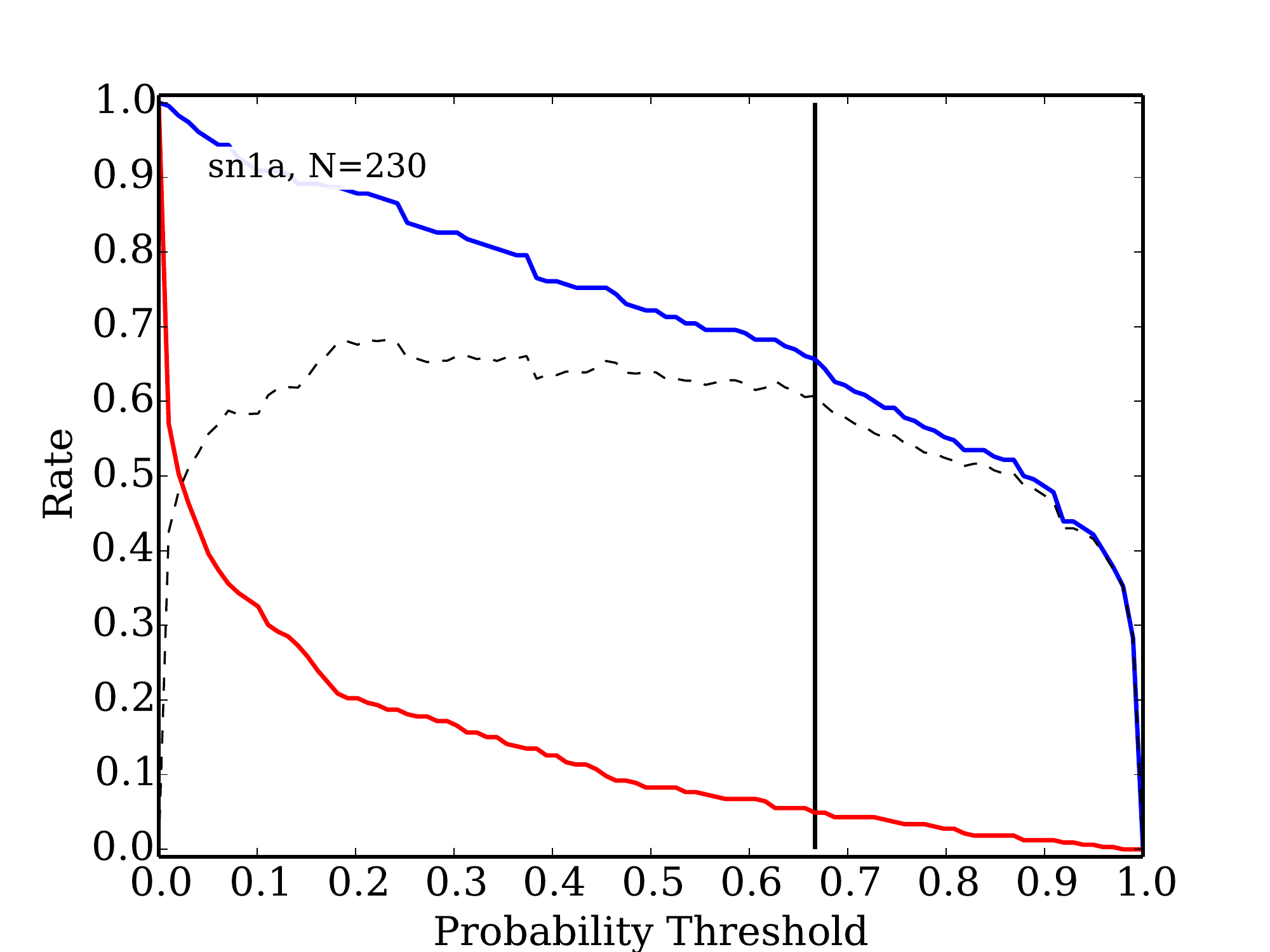} }}

}
\caption{Completeness$_{\text{SN Ia}}$ (blue) and Contamination$_{\text{SN Ia}}$ (red) as function of probability threshold for class SN Ia at magnitudes 16 ({\it top left}),17 ({\it top right}), 18 ({\it bottom left}) and 19 ({\it bottom right}). The vertical black line shows when the contamination is 5\%. The dashed line shows the completeness minus the contamination.}
\label{fig:contam_completeness}
\end{figure*}

\begin{table}
 \centering
  \begin{tabular}{@{}lllll@{}}
  \hline
  & M  & N & P(M) & \\ \hline
	& AGN 		& 500	& 0.053 	& \\
	& SN Ia 		& 5365	& 0.571 	& \\
    & SN Ibc 	& 295 	& 0.032 &  \\ 
    & SN II-P 	& 930	& 0.100 & \\ 
    & SN IIn 	& 400	& 0.043 & \\ 
    & Star 		& 1400	& 0.148 	& \\ 
    & Black body & 500	& 0.053 & \\ \hline
\end{tabular}
  \caption{N is the estimation of number of alerts of a given object type in the \textit{Gaia} survey for limiting magnitude $G$=19. P(M) is the prior probabilities for different object classes. See text for explanation.}
  \label{tab:rates}
\end{table}

The likelihood of the observed data $P(D, m_G, \vi| M)$ for a given model M is a likelihood marginalized over the parameters of the transient $\theta$ and is given by 
\begin{align}
P(D, m_G, \vi |  M)  = \int P(D, m_G, \vi| \theta, M) P(\theta|M) \ d\theta \label{eq:evidence} \\
P(D, m_G, \vi |  M) = \int P(D| \theta, M) P ( m_G| \vi, \theta, M) P ( \vi| \theta, M) P(\theta|M)  \ d\theta
\label{eq:prob_d_m_v}
\end{align}
The first component in the equation above is $P(D| \theta, M)$, (detailed in equation \ref{eq:likelihood}), and it is understood as the likelihood of the spectral data vector $D$ given the predicted vector $f(\theta,M)$ for model $M$ with parameters $\theta$. This is basically the product of individual Gaussian probabilities for each pixel position $i$. Note that both the observed $D$ and model spectra $f(\theta, M)$ are considered to be normalized, so they do not carry any information on the transient's magnitude. $P ( m_G| \vi, \theta, M)$ , in equation \ref{eq:magnitude_prob}, is the probability of seeing the transient at apparent magnitude $m_G$, given that the it belongs to class M with parameters $\theta$ and it is detectable by Gaia. $P(\vi|\theta, M)$ is the probability for the target to be visible and detectable by \gaia{} given its class and parameters (see equation \ref{eq:visibility}). Finally $P(\theta, M)$ is the joint prior on the models and its parameters.

\begin{equation}
P(D| \theta, M) = \prod_{i=1}^N P(d_i|\sigma_i, \theta, M)
\label{eq:likelihood}
\end{equation}
For simplicity we assume that the errors $e_i$ are uncorrelated and that the probability at each point $i$ only depends on the measured flux, $d_i$, estimated measurement error $e_i$ and the uncertainty of the true underlying model $\omega_i$. The likelihoods for every pixel $i$ will then be given by a Gaussian distribution with variance equal to the sum of the variance of data and the model. 
 \begin{align}
P(d_i | e_i, \theta, M) = \frac{1}{\sqrt{2 \pi ( e_i^2 + \omega_i^2) } } e^{-[d_i - f(i|\theta,M)]^2/2(e_i^2 + \omega_i^2) }
\end{align}

The choice on parameter prior function for each model, $P(\theta|M)$ is the prior on each one of the considered parameters: $P(z, t, A_V | M)$, based on observational constraints. Again for simplicity only three groups of priors have been considered: Supernovae (SN); extra-galactic objects (AGN); and stellar objects (STAR). The probability distribution of the reddening parameters for SNe spectra is computed as a third order polynomial based on the estimated $A_V$ values for all the training set derived during the standardization process. Redshift priors are proportional to the volume at that redshift up to a cut off value of $z=0.14$ for SNe and $z=1$ for AGN and galaxies, which represents the most distant expected transients that will be bright enough to de detected, i.e.
\begin{equation}
\text{P}(z) = 
\begin{cases} 
 \alpha \ z^2 + \beta 	& \text{if  } z \leq z_{lim} \\
0 						& \text{if  } z > z_{lim}
\end{cases}
\label{eq:z_prior}
\end{equation}
where $\alpha$ is a normalization constant and $\beta$ is a small constant (0.01) to allow for very low redshift events.

For the epoch of the SNe we use a uniform prior as in general we have no prior knowledge of when the SNe explosion occurred. This completely uninformative prior is a worst case scenario as during the \gaia{} mission there will be a non-detection history for the objects in addition to photometric information. This can be used to constrain the maximum epoch, or even distinguish if the transient is in pre-, or post-, maximum light phase.

\begin{figure*}
\centering
\mbox{
\subfigure{\includegraphics[width=\columnwidth]{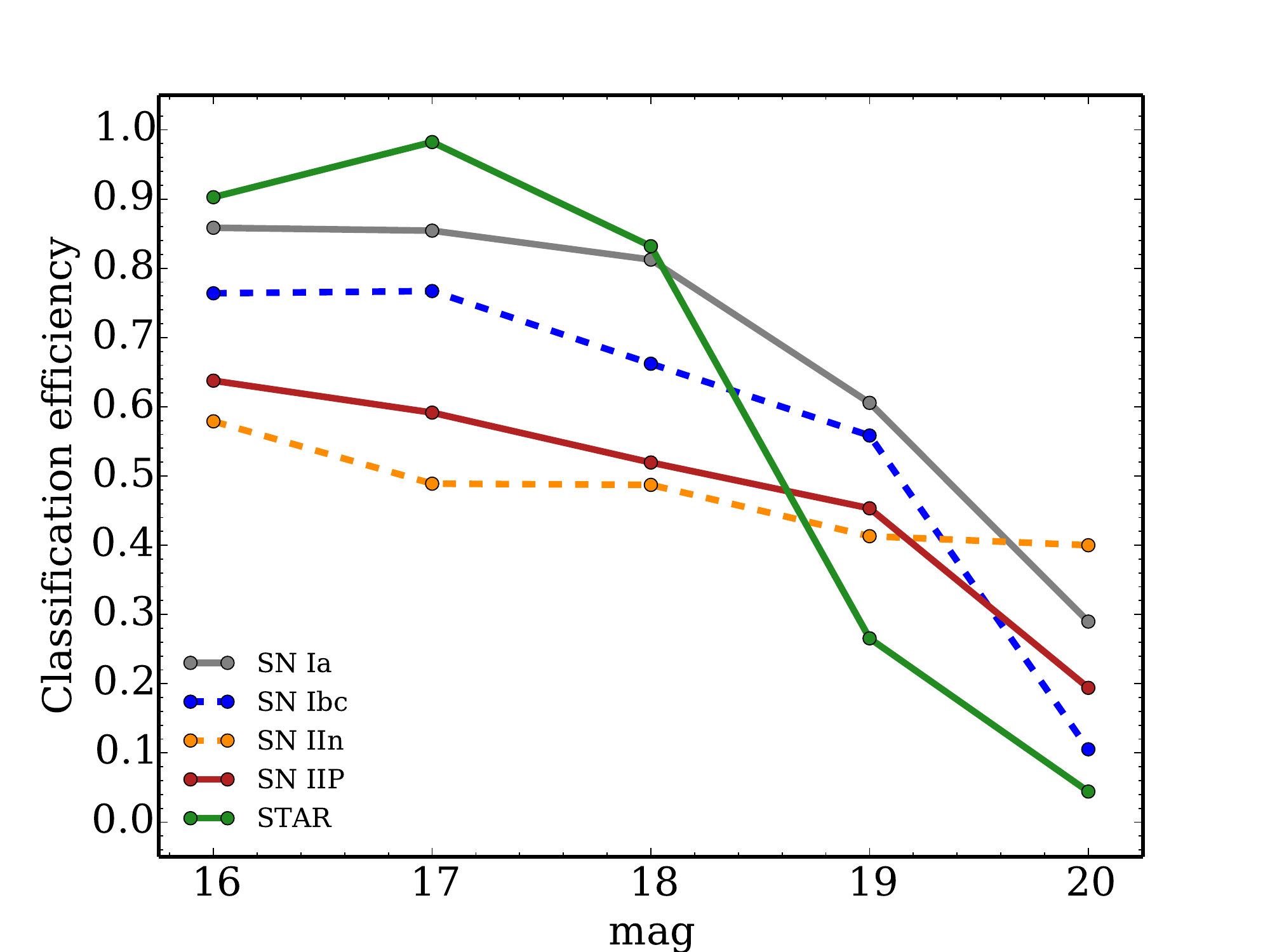} 
\quad
\subfigure{\includegraphics[width=\columnwidth]{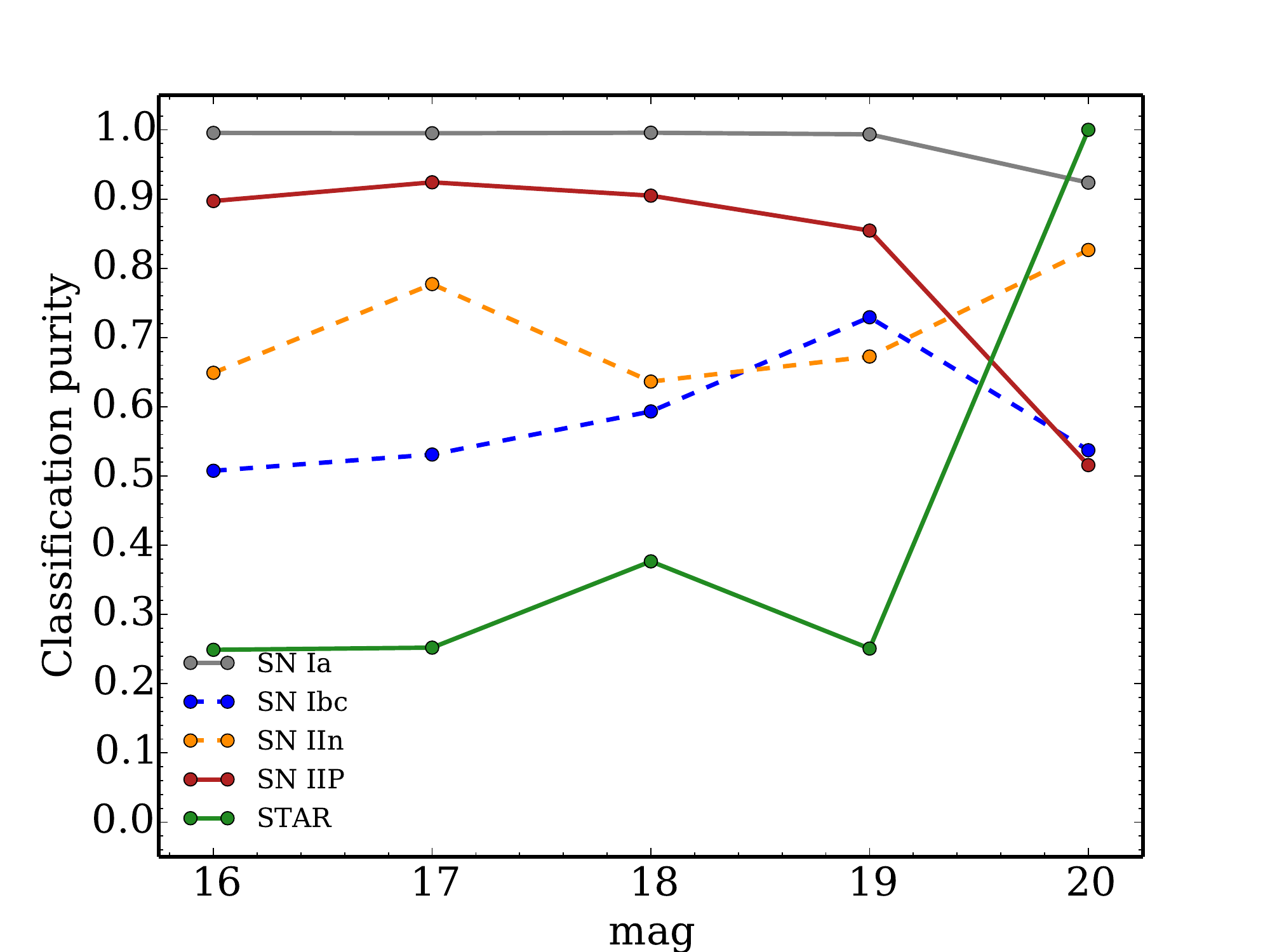} }} }
\mbox{
\subfigure{\includegraphics[width=\columnwidth]{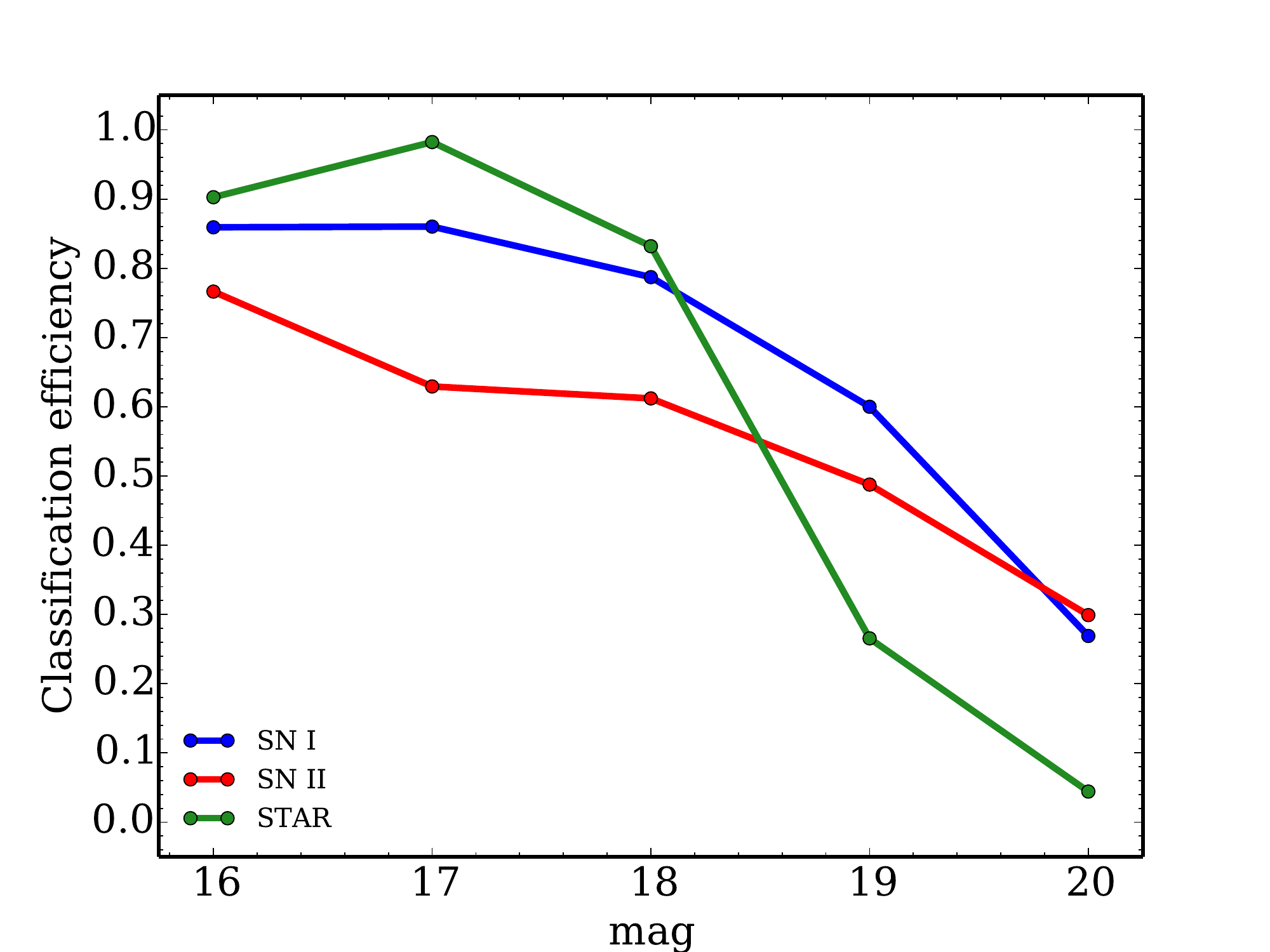}
\quad
\subfigure{\includegraphics[width=\columnwidth]{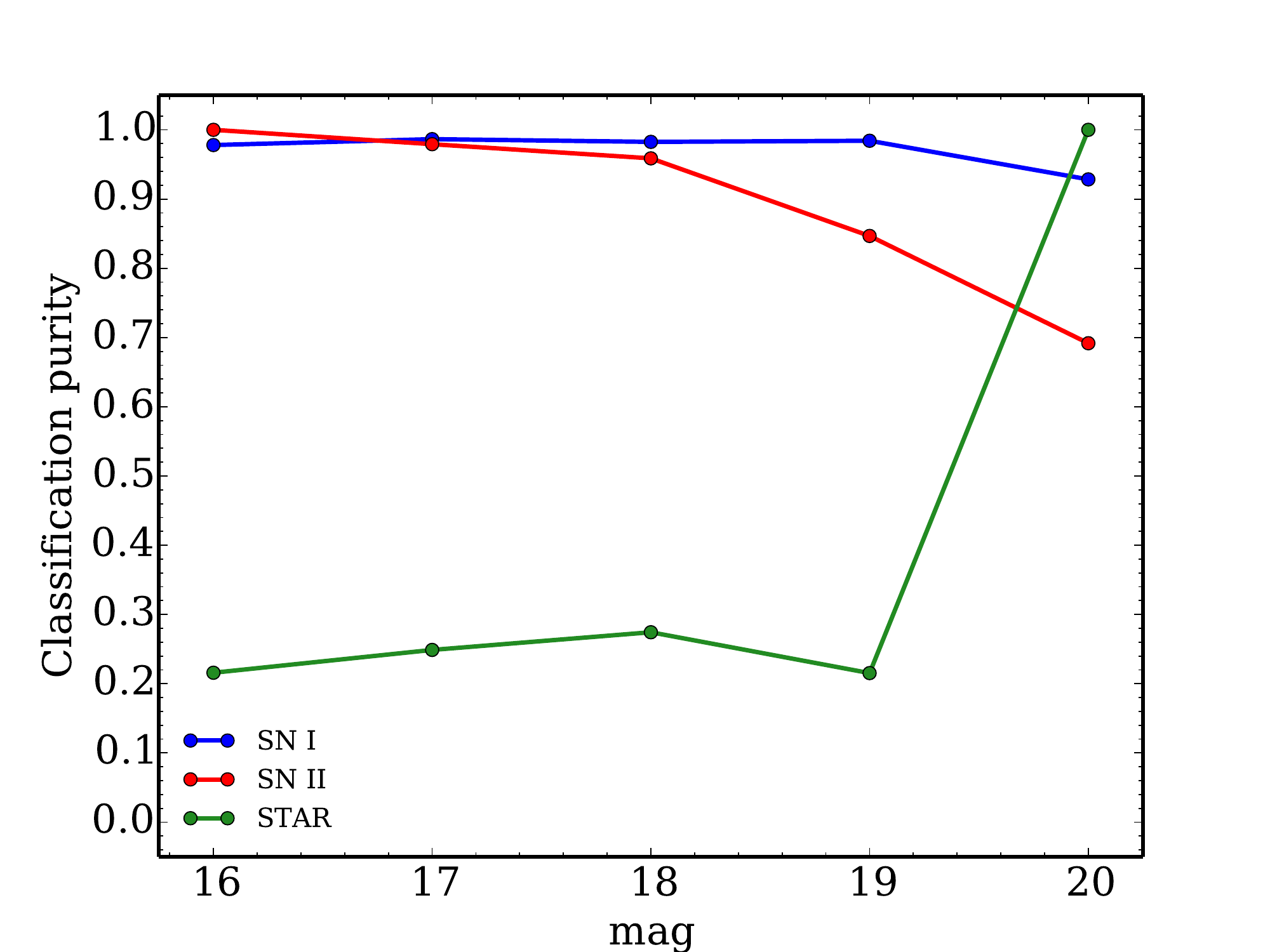} }}
}
\caption{{\it Top left}: Classification completeness (efficiency) as a function of magnitude for each individual class. {\it Top right}: Classification purity (e.g.  one minus contamination) as a function of magnitude and class. {\it Bottom left}: Classification completeness as a function of magnitude for the two main SNe subtypes. {\it Bottom right}: Classification purity as a function of magnitude and class for the two main SNe subtypes.}
\label{fig:accuracy}
\end{figure*}

Using additional information and prior knowledge of the behaviour of objects to aid classification is common practice \citep{Bailer-Jones2011}. It is particularly  useful for discarding parts of the parameter space which can not be populated because of physical constraints. Information on the object apparent magnitude and the fact that the event was detected by \gaia{} can be used to significantly reduce the prior space for $\theta$. Following this approach we introduce the object's visibility \vi and apparent magnitude in the \gaia{} $G$ band, $m_G$. Then, given a class and a parameter space, we can compute how likely the objects are to be visible by \gaia{} for each point in this space, and how likely each point will be seen at apparent magnitude $m_G$. To compute this information for each model we need the following ingredients:

\begin{itemize}
 \item an absolute magnitude at the epoch of maximum brightness in the $V$ band \citep{Li2011}. As the visual $V$ band is close to the \gaia{} $G$ band for blue transients \citep{Jordi2010}, the peak absolute magnitude for class M can be described by a normal distribution of mean $G_M$ and standard deviation $\sigma_M$i.e. $\mathcal{N}(G_M, \sigma_M)$. In order to make our priors less dependent on statistical fluctuations, increasing our ignorance on the true underlying absolute magnitude distribution, the adopted standard deviation is twice the value provided by literature;
 \item the evolution of the object's magnitude as a function of the epoch relative to maximum brightness, where the brightness at each epoch is modelled as a Gaussian distribution around the mean light curve $m_{lc}(t)$ with standard deviation $\sigma_{lc}(t)$ i.e. $\mathcal{N}(m_{lc}(t), \sigma_{lc}(t))$. The light curves and dispersions of the light curve for each model have been computed from \cite{2011yCat..74121441L};
 \item $d$ the luminosity distance at redshift $z$ assuming the cosmological parameters \{$\Omega_M$=0.3, $\Omega_{\Lambda}$= 0.7, $h$=0.7\}.
\end{itemize}
Overall, the theoretically predicted apparent magnitude, $m_T(\theta, M)$ is a combination of the object absolute peak magnitude, light curve phase, luminosity distance and the amount of extinction along the line of sight. 

\begin{align}
m_T(\theta|M) = G_M + 5 \text{log}(d) - 5 + A_V + m_{lc}(t) \\
\sigma_T(\theta|M) = \sqrt{\sigma_{M}^2 + \sigma_{lc}(t)^2}
\end{align}

 The probability that the transient is seen at an apparent magnitude $m_G$ only holds when $m_G \leq m_{lim}$.
\begin{align}
P(m_G|\vi, \theta,M) = \frac{P(m_G|\theta,M)}{P(\vi|\theta, M)} \\
P(m_G|\theta,M) = 
\frac{1}{\sqrt{2 \pi \sigma_T(\theta|M)^2  } }
\text{exp} 
\left\{  
- \frac{\left[ m_G - m_T(\theta|M) \right] ^2  }{2(\sigma_T(\theta|M)^2)}
\right\}
\label{eq:magnitude_prob}
\end{align}

In order to obtain a normalized distribution, we should normalize to the object visibility: $P(\vi | \theta, M)$ which is the third component in eq.~\ref{eq:prob_d_m_v}. It accounts for discarding all the parameter space for model M where objects  would be too faint to be detected by the satellite, or in other words, $m_T > m_{lim}$. Therefore, the probability that the object is detectable by the satellite is the cumulative probability distribution for an object of type M with parameters $\theta$ to be brighter than $m_{lim}$,
\begin{align}
P(\vi| \theta, M)= \frac{1}{\sqrt{2 \pi \sigma_T(\theta|M)^2  } }
\int_{-\infty}^{m_{lim}} 
\text{exp} \left\{  - \frac{ \left[ m - m_T(\theta|M) \right] ^2} {2\sigma_T(\theta|M)^2}
\right\} \text{dm}
\label{eq:visibility}
\end{align}

\subsection{Assessing classification performance} \label{sec:model_selection}

 In general the classification
result may be ambiguous with several models having similar probabilities for a given sample of input data. The reasons could be diverse ranging from low signal-to-noise in the input data to unforeseen types of transient. We aim to identify these cases and mark them as \textit{ambiguous} classifications. This means that only classifications considered to be reliable can be selected, leaving characterization of uncertain targets to further observational follow-up.

The key parameters of the classifier performance are the completeness and contamination. Following the definition used in \cite{Bailer-Jones2008} we define the classification and completeness for each class $j$ as
\begin{align}
\text{Completeness}_j = \frac{n_{k=j, j}}{N_k} \label{eq:complete}
\\
\text{Contamination}_j = \frac{ \sum_k n_{k \neq j,j} }{\sum_k{n_{k,j}} } 
\label{eq:contam}
\end{align}
where $n_{k,j}$ is the number of objects of true class $k$ classified as class $j$, $N_k$ is the total number of objects of class $k$ in the test set, $n_{k=j, j}$ refers to correctly classified objects, $n_{k \neq j, j}$ includes all misclassified objects for class $j$ and $\sum_k{n_{k,j}}$ is the total number of objects classified (correctly or not) as class $j$. 

These parameters are used to help select the optimum probability threshold for each class to achieve high robustness in the classification results.  However, as
the relative fractions of objects of different classes in the training set do not match the prior fractions in Table~\ref{tab:rates} we adjust the calculation 
of contamination using weights reflecting the relative frequency of class $k$ over class $j$ in the training set, $f_{k/j}^{train}$ and the expected fraction during mission: $f_{k/j}^{real}$.
 \begin{align}
\text{Contamination}_j^w = \frac{ \sum_{k \neq j} \  n_{k,j} (f_{k/j}^{real} / f_{k/j}^{train}) }{\sum_k{n_{k,j} (f_{k/j}^{real} / f_{k/j}^{train})} } \label{eq:contamination_weighted} 
\end{align}
It is then useful to introduce the concept of Purity since it defines how reliable the classification is once we have provided a reliable answer
 \begin{align}
\text{Purity}_j^w = 1 -  \text{Contamination}_j^w \label{eq:purity} 
\end{align}

\subsection{Parameter estimation} \label{sec:param_estim}

For some objects we will be interested not only in their class, but also in their other properties, such as redshifts and epochs. To determine these we use
the posterior probability distributions for the parameters of interest (redshift and epoch in our case), which are in turn obtained by marginalizing over the remaining parameters:

\begin{align}
P(z | D, M) = \int_{t} \int_{A_V} P(D, m_G, \vi|t, z, A_V, M) P(\theta | M) \ dt \ dA_V  \label{eq:marg1}\\
P(t | D, M) = \int_{z} \int_{A_V} P(D, m_G, \vi|t, z, A_V, M) P(\theta | M) \ dz \ dA_V  \label{eq:marg2}
\end{align}

The peak of the marginalized probability distribution is the most likely value for the parameter of interest. To estimate the 1$\sigma$ errorbars on a (generally) non-Gaussian distribution is to make the assumption that near the peak, the probability behaves as a Gaussian. In this case we fit a second degree polynomial to the logarithm of the probability distribution around the peak to provide an estimate of $\sigma$ \citep{Lampton1976}.

\section{Transient classification results}  \label{sec:results}

\subsection{Test Configuration description} \label{sec:test_set}

The verification of our classification algorithm was done using a K-fold cross-validation, with $K=10$. The total set of spectra is divided into 10 randomly selected non-overlapping sets, each one containing 10\% of the total sample. For each set the classifier is trained with the remaining 90\% of the spectra such that the same spectra are never used together for testing and training. The results of all 10 sets are then combined into a single set for analysis purposes.

The main aim of this test is to assess the effect of magnitude (primarily signal-to-noise) on the classification accuracy. However, as  the original magnitudes of the objects were not available in the spectral archives, we estimated them from the object redshifts, epochs and extinction values computed during the standardization process described in Section~\ref{sec:standard}.
The resulting data set, as expected, contained very few spectra with bright magnitudes and low redshift and many more spectra with faint magnitudes at higher redshift. In order to robustly assess how well the classifier performs at different magnitudes with the same set of spectra, we artificially shift the spectra to higher or lower redshifts. By varying the distance modulus of the object we can uniformly populate the magnitude bins from 16 to 20. Tests with objects brighter than $G$=16 have shown similar characteristics to those with magnitude 16 and therefore have been omitted. 
  
The computing performance of the algorithm developed in the pipeline is on average between 5 and 6 s per classification object on a single core Intel (R) Core (TM) i7-2600 CPU 3.4GHz, which makes it suitable for analysing around 10 000 transient alerts in less than 2 hours in an 8 cored workstation.

\subsection{Classification Accuracy} \label{sec:accuracy}

\begin{figure*}
\centering
  \subfigure{\includegraphics[width=1.6\columnwidth]{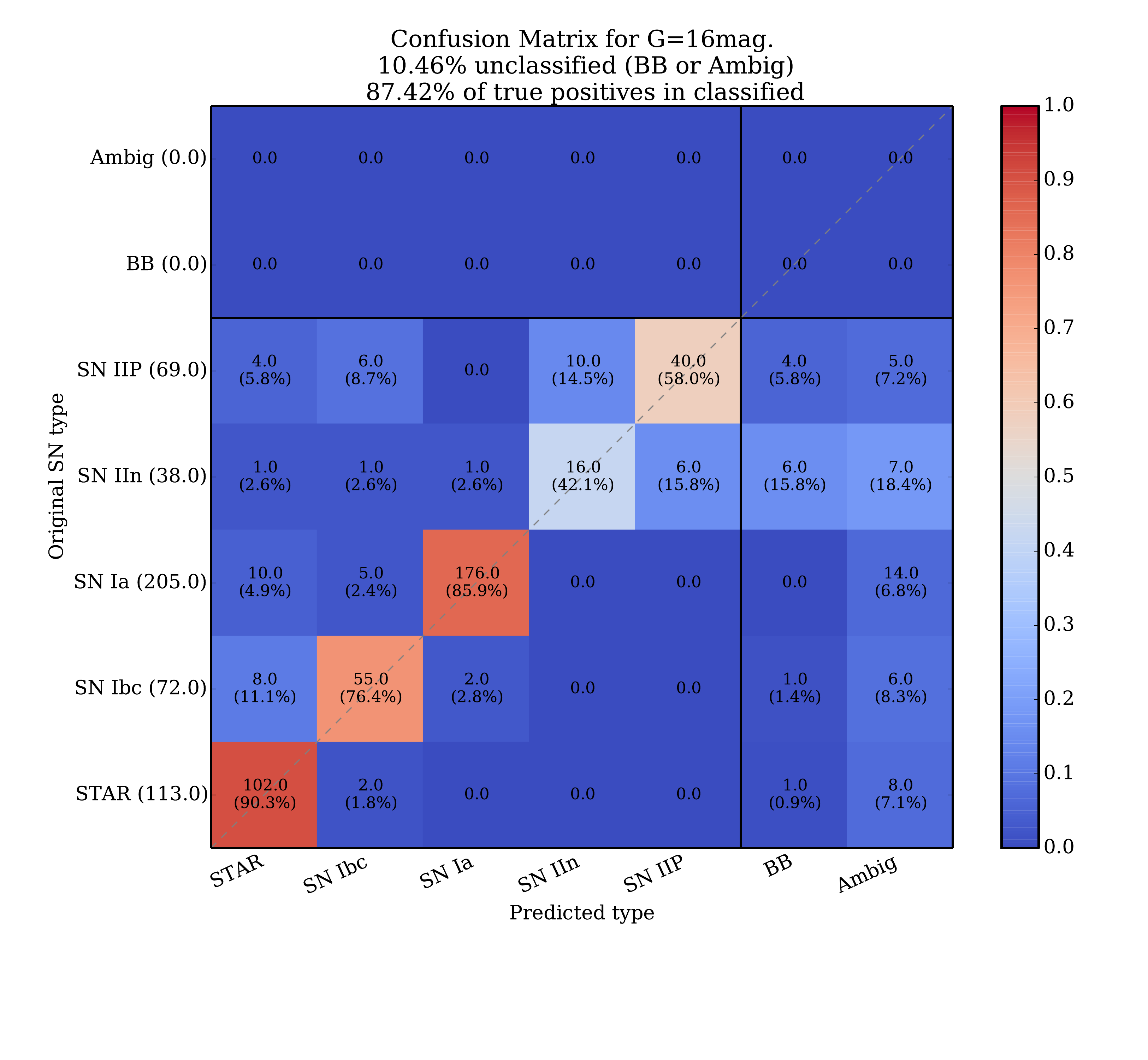}}\\ \vspace{-7em}%
  \subfigure{\includegraphics[width=1.6\columnwidth]{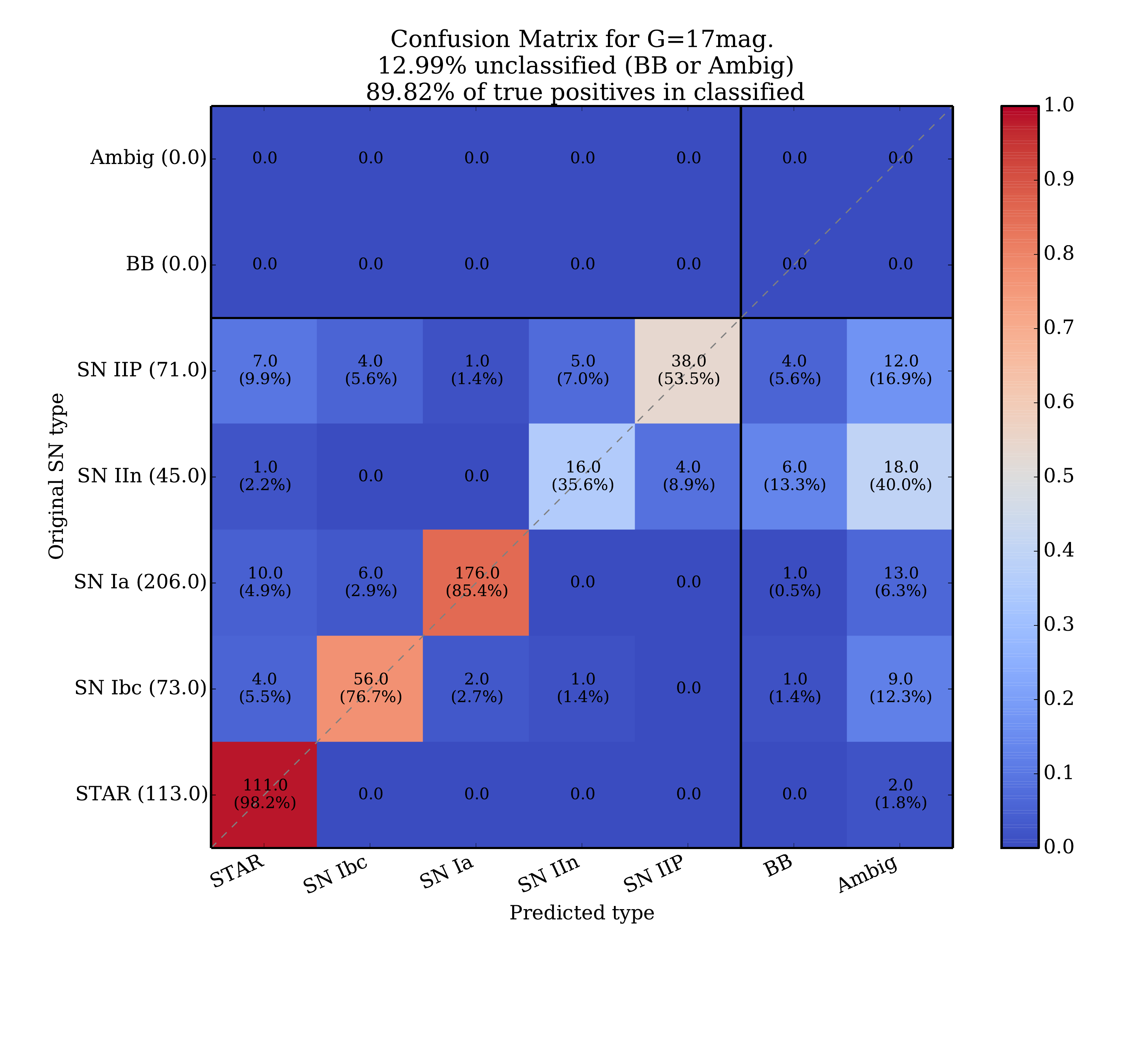}}
  \vspace{-7em}
      \caption{Confusion matrices for the bright end. The X axis represents the class type predicted by the classifier and the Y axis represents the true type. The number in parenthesis is the number of spectra used in the test set. The percentages are given relative to this number. The black line separates the real types from the artificial types: BB and Ambiguous. The color bar indicates the percentage of objects that belong to each category.}
\label{fig:confumat_bright}
\end{figure*}

\begin{figure*}
\centering
  \subfigure{\includegraphics[width=1.6\columnwidth]{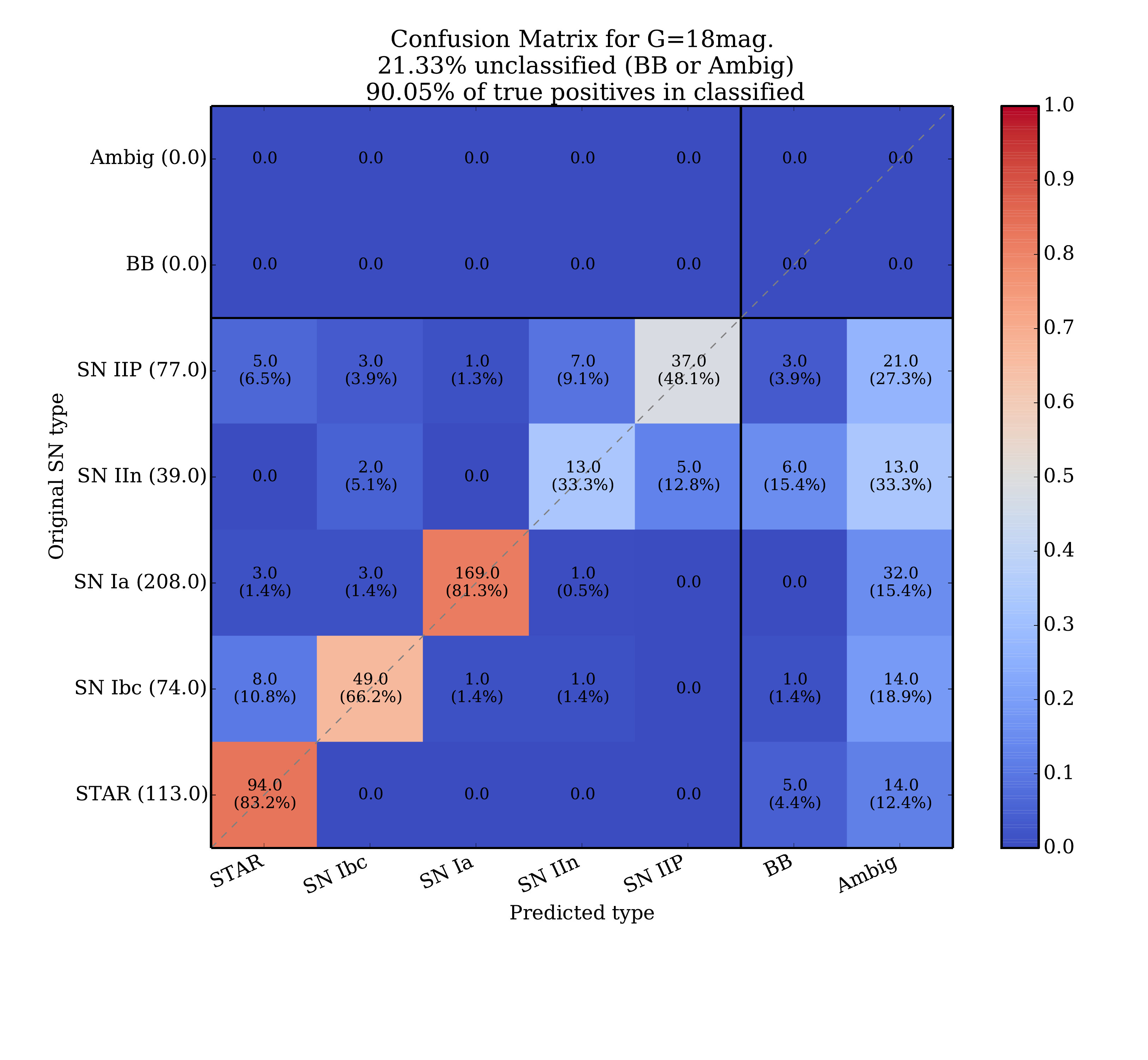}}\\ \vspace{-7em}%
  \subfigure{\includegraphics[width=1.6\columnwidth]{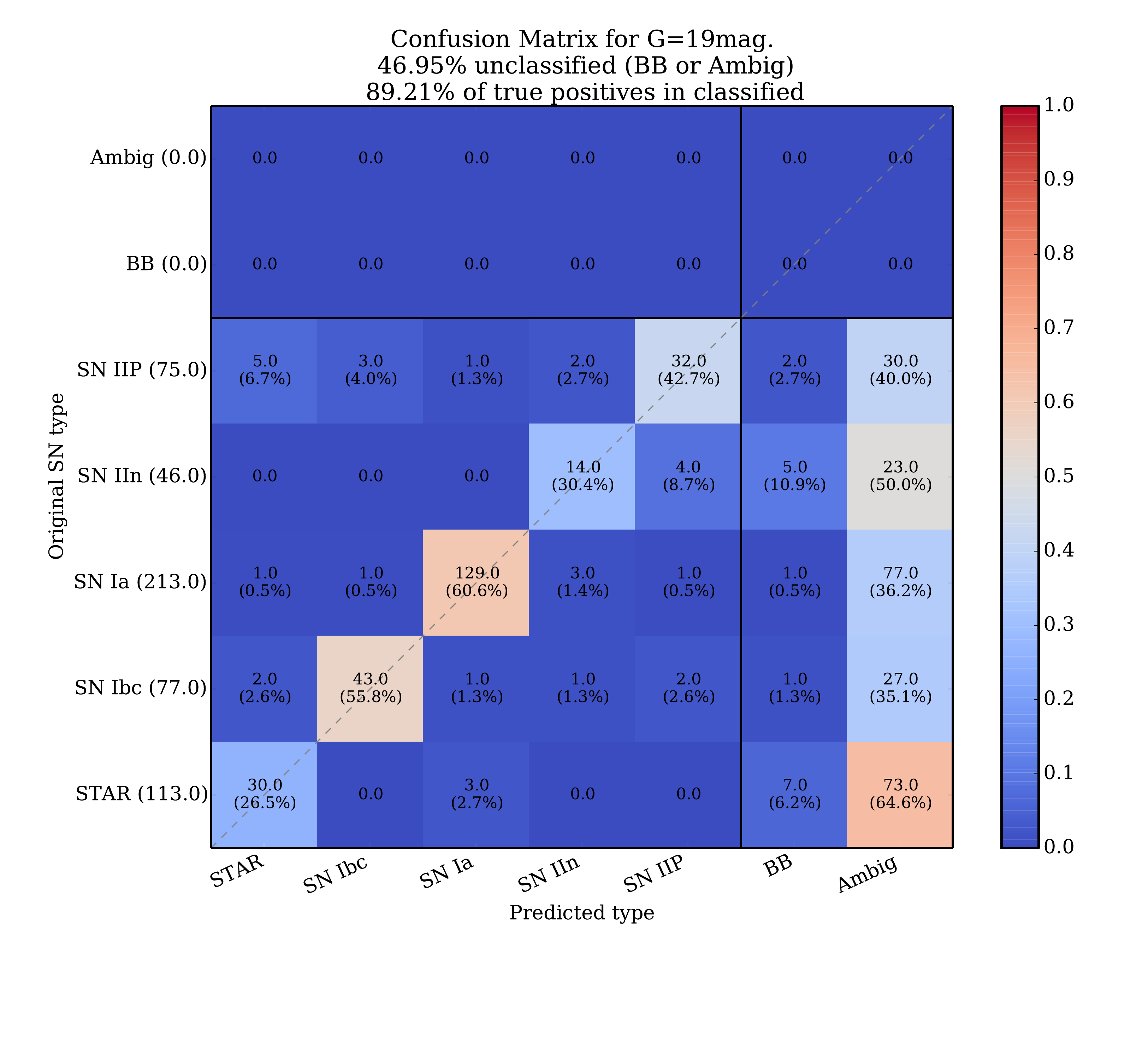}}
    \vspace{-7em}
    \caption{Confusion matrices for the faint end. The X axis represents the class type predicted by the classifier and the Y axis represents the true type. The number in parenthesis is the number of spectra used in the test set. The percentages are given relative to this number. The black line separates the real types from the artificial types: BB and Ambiguous. The color bar indicates the percentage of objects that belong to each category.}
    \label{fig:confumat_faint}
\end{figure*}

This section presents the results of our tests with magnitudes ranging from $G$=16 to $G$=20. As noted in section \ref{sec:model_selection}, the classification performance is assessed by two separate metrics: \textit{classification completeness}; and \textit{classification contamination}. By observing how these metrics behave for each class and magnitude we can choose the optimum class probability thresholds, $p$, where the classification is considered to be \textit{reliable}. Transients with probabilities lower than the threshold are marked as \textit{ambiguous} cases and therefore left unclassified; transients with higher probabilities are considered \textit{reliable} classifications.

Figure \ref{fig:contam_completeness} shows an example of this analysis for SNe type Ia as a function of the class probability. The blue line represents the completeness of the sample defined in equation \ref{eq:complete} and the red line the class contamination, defined in equation \ref{eq:contam}. Brighter magnitudes have high completeness and low contamination levels, even at low probability thresholds, due to the objects having good signal-to-noise. For fainter objects misclassification increases as more classes start to resemble the noisy spectra.

In our case we want to keep the contamination low and consequently use a conservative probability threshold will introduce less than 5\% of contamination. Figure \ref{fig:contam_completeness} (unsurprisingly) demonstrates that contamination increases significantly with magnitude and forces us to adopt an increasing probability threshold for SNe type Ia from 0.5 at magnitude 16 to 0.9 at magnitude 20. For the remaining types of SNe, a selected threshold of 0.3 is enough to keep the contamination below 5\%. The selection threshold is different for variable stars, as they are the biggest source of contamination for SNe at bright magnitudes. Therefore, a higher threshold of 0.9 is used to select variable stars at magnitude 16, but only 0.5 on faint magnitudes, when they no longer resemble SNe. We discuss this contamination issue further ahead in this section.

Figure \ref{fig:accuracy} shows the completeness of the classification and its purity for each object type and magnitude. Completeness decreases as objects become fainter and more of them become classified as \textit{ambiguous} sources. We observe that for bright sources at magnitude 16, almost all object types can be identified with an efficiency of 70 to 90 \%, as shown in Table \ref{tab:purity}.

Fainter objects generally have too low a signal-to-noise ratio to give a reliable classification solely from the spectral shape, the $P(D, m_G, \vi |\theta, M)$ data component in equation \ref{eq:model_prob}.  In these cases the prior probability on the object model type $P(M)$ becomes dominant. In the faint magnitude regime, objects will generally only be reliably classified if they are a very good match with the library objects and also have a high prior probability.

The purity of the classification is strongly dependant on the object type. For the most common type, SNe Ia, the purity is around 99\% for almost all the magnitudes, as shown in Table \ref{tab:purity}. Any contamination coming from less frequent classes has little effect.  We see the opposite for less common types, such as SNe Ibc which spectrally resemble SNe Ia at early epochs \citep{Filippenko1997}. This class type may accidentally receive the label of SN Ia and given the latter's high frequency, the purity for SNe type Ibc will be considerably lower.

For bright magnitudes some SNe Ia are confused with variable stars. This is shown in the confusion matrices for the brighter magnitudes 16 and 17 in Figure \ref{fig:confumat_bright}. This is due to objects with very weak features, such as highly reddened SNe, or spectra with strong host galaxy component. When these objects are at very low redshifts they can look like variable stars.

SNe of magnitude 16 and 17 must be very nearby and according to our priors, at such magnitudes variable stars are much more likely than SNe, therefore a slightly lower efficiency for bright SNe and a decreased purity for bright variable stars is expected. However, this stops being an issue at magnitudes 18 and higher, as spectra for fainter SNe are noticeably redshifted becoming more distinguishable from the spectra of variable stars and therefore reducing the misclassification rate.

At very faint magnitudes the information contained in the detailed spectral shape is less dominant so that it becomes harder for a given type to score above the probability threshold. In this regime transients can be fit by black body spectra or other alternate types. This effect can be observed in the confusion matrices for fainter magnitudes in Figure~\ref{fig:confumat_faint} which show larger number of objects labelled as BB (black body) or Ambig (ambiguous).

At early epochs SNe type IIn usually have weak and narrow \ha emission which is barely visible in the BP/RP spectrum.  These objects are often classified as black bodies as they generally lack other major spectral features. For this reason, in the current work, for both SNe type IIP and SNe type IIn at epochs younger than 5 days, a BB type was adopted as a valid answer. 

For later epochs SNe type IIn are well classified, even at faint magnitudes. These SNe generally develop very strong emission in \ha. This line is well mapped by the red part of the spectrograph, which also has slightly higher resolution than blue spectrograph, and therefore this line can be still be identified even at low signal-to-noise. 

Purity can decrease at fainter magnitudes for SNe type IIP since they resemble SNe IIn with both having strong \ha and Ca II emission lines. In the low signal-to-noise regime the characteristic wide p-Cygni profile of SNe IIP is not well recognized, especially if the lines are relatively weak. Truncation of the test spectra can also be a source of confusion among SNe IIP and SNe IIn, as the Ca II emission line may be totally missing.

Due to the confusion of detailed spectral types mentioned above we took the decision to offer a more general classification answer, where we merge similar SNe subtypes and offer classification labels such as SN I, SN II, STAR, AGN or BB, with higher reliability. We present parameter estimates for these general types as well.

 \begin{figure}

\hspace{-0.5cm}
\subfigure{\includegraphics[width=1.15\columnwidth]{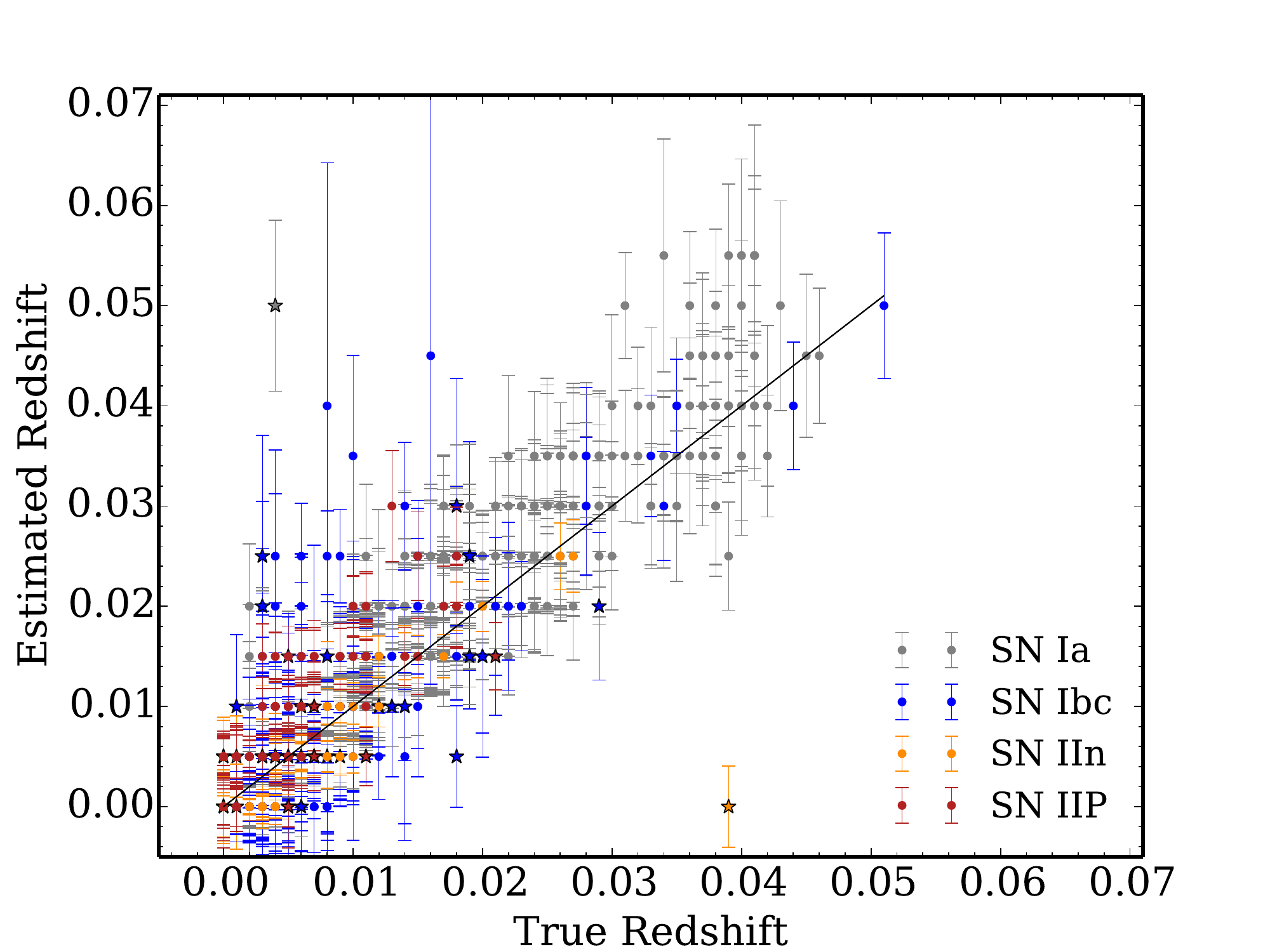}}
\quad

\subfigure{ \hspace{-0.5cm}\includegraphics[width=1.15\columnwidth]{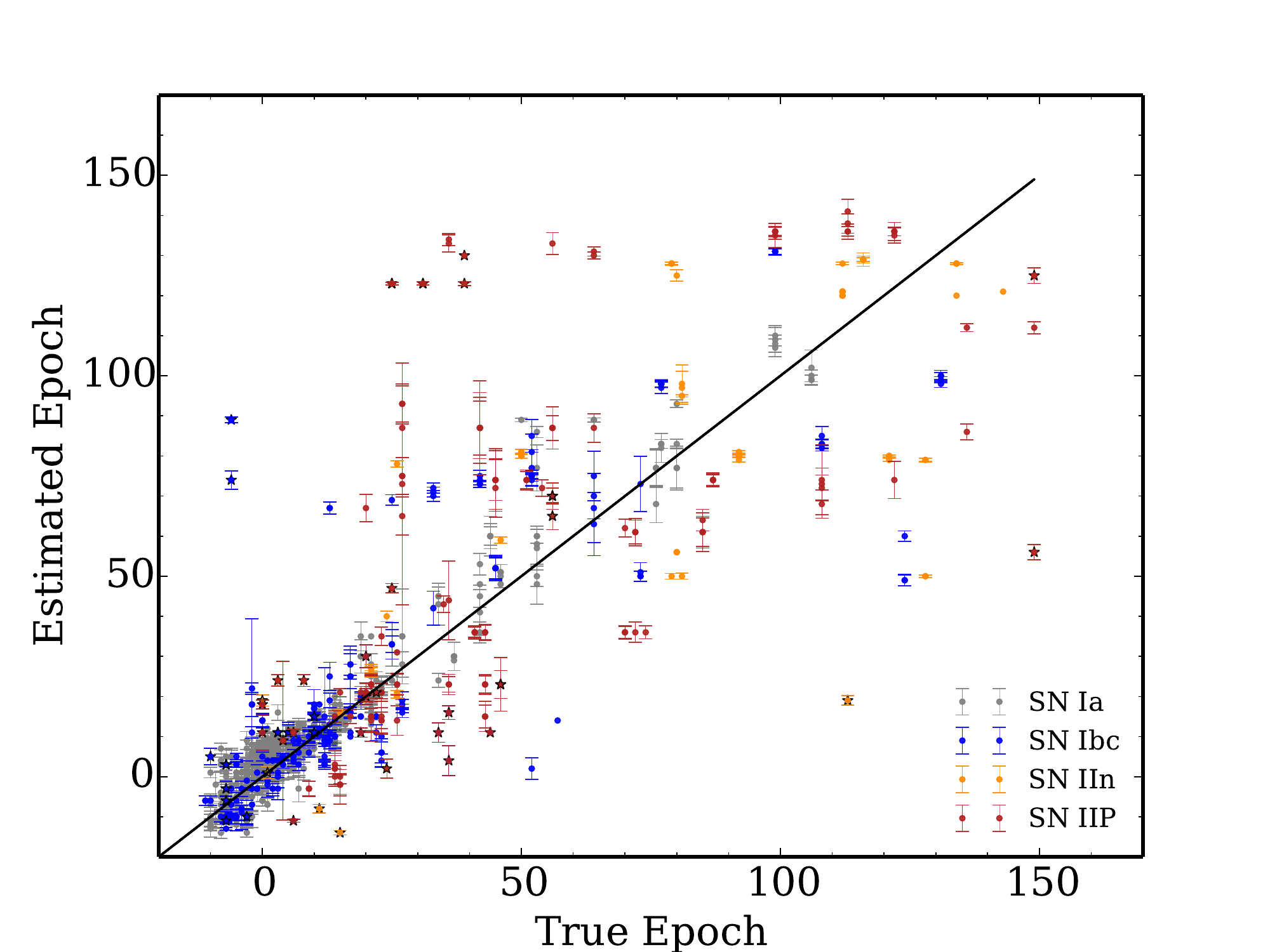} }

\caption{{\it Top}: Scatter plot that shows the performance of redshift parameter estimation. Estimated values for redshift (Y axis) are plotted against the true values from the spectral archive (X axis). Star symbols represent false positives for each labelled class. {\it Bottom}: Scatter plot that shows the performance of epoch parameter estimation. Estimated values for the object epoch (Y axis) are plotted against the true epoch values from the spectral archive (X axis). Star symbols represent false positives for each labelled class.
}
\label{fig:parameters}
\end{figure}

\subsection{Parameter Estimation Accuracy} \label{sec:param_accuracy}

\begin{table*}
 \mbox{
  \begin{tabular}{@{}llllllll@{}}
  \hline
  \multicolumn{8}{l}{ \textit{Classification efficiency}} \\\hline
Mag  & SN Ia & SN Ibc & SN IIn & SN IIP & SN I & SN II & STAR  \\ \hline
16.0 &85.9 &76.4 &57.9 &63.8 &85.9 &76.6 &90.3 \\
17.0 &85.4 &76.7 &48.9 &59.2 &86.0 &62.9 &98.2 \\
18.0 &81.2 &66.2 &48.7 &51.9 &78.7 &61.2 &83.2 \\
19.0 &60.6 &55.8 &41.3 &45.3 &60.0 &48.8 &26.5\\
20.0 &29.0 &10.5 &40.0 &19.4 &26.9 &29.9 &4.4 \\
\hline
\end{tabular}
\hspace{0.1cm}
  \begin{tabular}{@{}llllllll@{}}
  \hline
    \multicolumn{8}{l}{ \textit{Classification purity}} \\\hline
    
Mag  & SN Ia & SN Ibc & SN IIn & SN IIP & SN I & SN II & STAR  \\ \hline
16.0 &99.6 &50.8 &64.9 &89.7 &97.8 &100.0 &21.6\\
17.0 &99.5 &53.1 &77.7 &92.4 &98.7 &97.9 &24.9\\
18.0 &99.6 &59.3 &63.6 &90.5 &98.3 &95.9 &27.4\\
19.0 &99.3 &72.9 &67.3 &85.4 &98.4 &84.7 &21.5\\
20.0 &92.4 &53.7 &82.6 &51.6 &92.8 &69.2 &100.0\\
    \hline
\end{tabular}

}
\caption{Percentage classification efficiency and weighted purity for the cross-validation tests. Values are provided for individual types and for joined SNe type I and SNe type II.}
 \label{tab:purity}
\end{table*}

The parameter estimation accuracy is computed for each transient class for all transients brighter than 19.5 magnitude. The results for redshift and epoch determination are displayed in Figure \ref{fig:parameters}. The top plots show the true (X axis) and the estimated (Y axis) parameters for each individual object. Dots represent correctly classified objects and stars refer to false positives for each class.

There are two noticeable effects in the redshift scatter diagram. First, as expected, we see that for nearby objects the scatter is lower since these objects generally are brighter and hence have better signal-to-noise. Second, low redshift objects are more likely to be misclassified because of the confusion with variable stars explained previously.

The scatter plot for epoch determination of epochs in Figure~\ref{fig:parameters} also shows lower dispersion for early epochs. This is also as expected since young objects normally evolve quickly allowing a tighter constraint on their epoch. 

Redshift is predicted with an accuracy of $\sigma_{z} \simeq 0.008$ for SNe type II and $\sigma_z \simeq 0.006$ for SNe type I. The average error on epoch is $\sigma_{t} \lesssim 13$ days for SNe type I, and around 31 days for SNe type II. If we restrict selection to objects with epochs younger than 50 days post maximum-light, the epoch scatter reduces to 8 and 30 days for SNe type I and II respectively.

The outliers in redshift determination that appear in Figure \ref{fig:parameters} are generally misclassified objects. These are clearly a minority as evinced by the error histogram distributions.

\subsection{Accuracy with improved S/N} \label{sec:improved}

\begin{figure}
\hspace{-0.7cm}
\includegraphics[width=1.15\columnwidth]{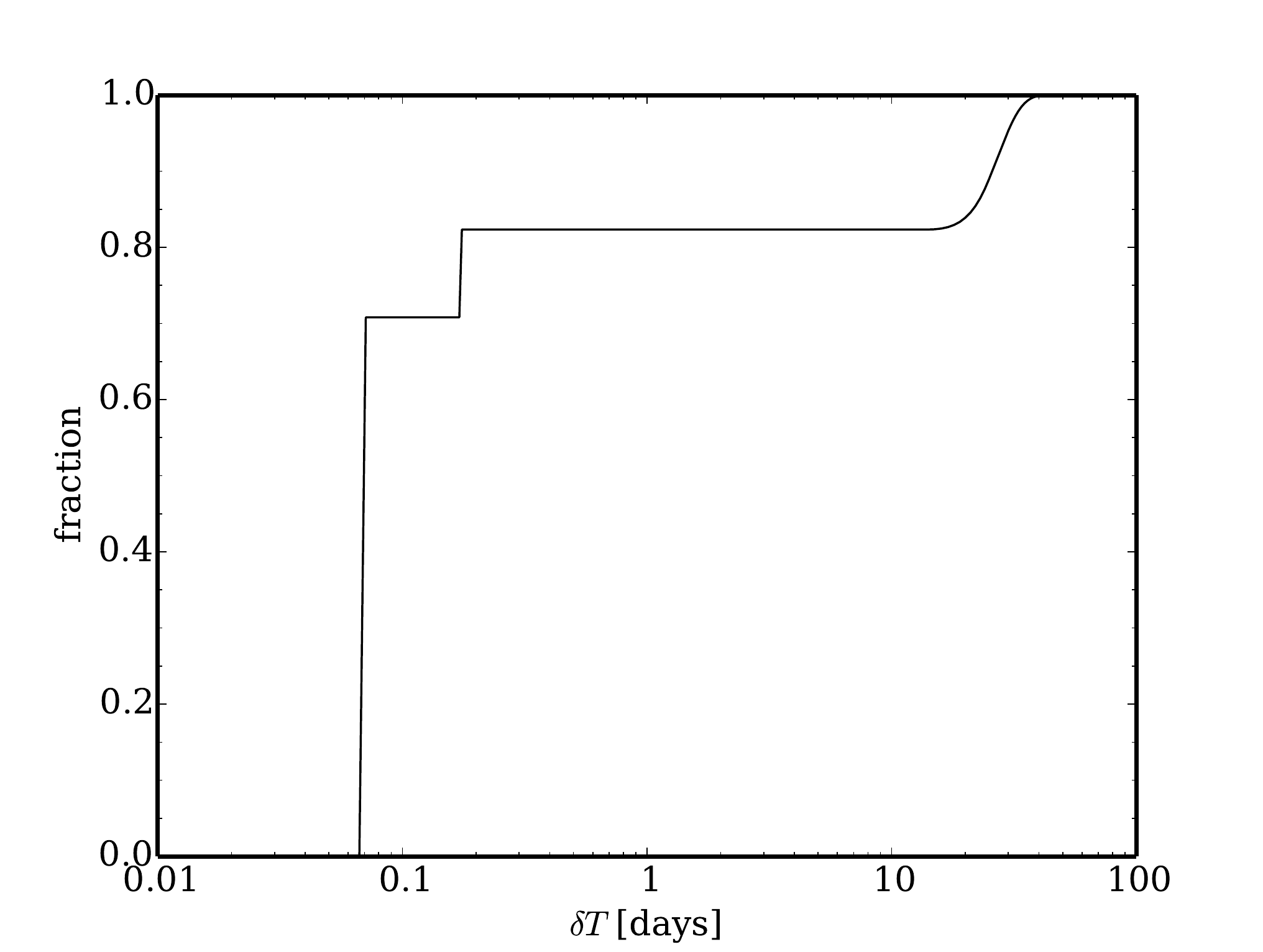} 
\caption{Probability of having a repeated \gaia{} observation on the same field within $\delta T$ days. Around 70\% of the fields will be observed at least twice with an interval of 106 min. Around 80\% will be re-observed in less than one day. The final fraction heavily increases after an interval of 30 days.}
\label{fig:reobs}
\end{figure}

The results of the classification process are computed for single transients. However, as shown in Figure \ref{fig:reobs}, around 70\% of sky will have a second observation very close in time, generally 106 minutes later. An additional 10\% will have another observation around 4h later. The data will be taken under similar conditions, instrument configuration and scanning angle, therefore it is reasonable to test how much the performance improves if we use stacked spectra in order to improve the signal-to-noise ratio.

Tests with two stacked spectra show that although there is a positive effect, this is quite small. 
Stacking has a slight positive effect on the classification results: the efficiency increases between 5\% and 10\% for SN I and SN II, but it decreases for stars in the faint end by around 10\%. The purity generally improves for both SNe and stars, specially at the faint end. However, this  effect is always below 10\%.

The conclusion from this test is that while at brighter magnitudes stacking several spectra is not ultimately improving the results, for fainter objects which are going to be the majority, the stacking process may improve the completeness and purity of the classification results.
The improvement in parameter determination is practically negligible.

%

\section{Application to PESSTO data}\label{sec:pessto}
In the previous sections we described the Bayesian forward method for \gaia{} SNe classification. We have also shown how the method performs on the test set that we constructed. In this Section we apply this method to the PESSTO transient dataset. This test is important, because it constitutes a separate validation test, carried on data coming from a single source (EFOSC2 instrument on the NTT telescope in La Silla). 

One of the main goals of PESSTO is the identification of non-standard transient objects \citep{Valenti2014}. Therefore, continuous classification of new transients has been performed by the PESSTO team since April 2012 (\cite{PESSTO2012}, \cite{ESO_Messenger}). 1117 transients have been made available so far via the WiseRep spectral data repository. However, we had to discard almost half of these because of insufficient wavelength coverage together with uncertain classification labels, such as \textit{Other} and \textit{Unknown}, leaving 507 spectra. Unfortunately, many objects have spectra taken at only a single epoch and have no accurate estimate of the epoch parameter. Apparent magnitude in the optical bands is often missing as well, as the photometry is done separately (for follow-up objects), or is not done at all (for those classified objects that have not been selected). As the apparent magnitude is an important parameter for GS-TEC we have estimated it from the flux measurements of spectra with broad wavelength coverage using the \texttt{python} package \texttt{pysynphot} and the \gaia{} G-band response.

The medium-resolution spectra from PESSTO were converted to \gaia{} low-resolution spectra using these estimated magnitude values. As some of the PESSTO targets were observed when their magnitudes were too faint for \gaia{}, we applied a small blueshift correction such that they could be included in the sample for testing purposes.

\subsection{Results for PESSTO data}

\begin{figure*}
\centering
\mbox{
\subfigure{\includegraphics[width=1.1\columnwidth]{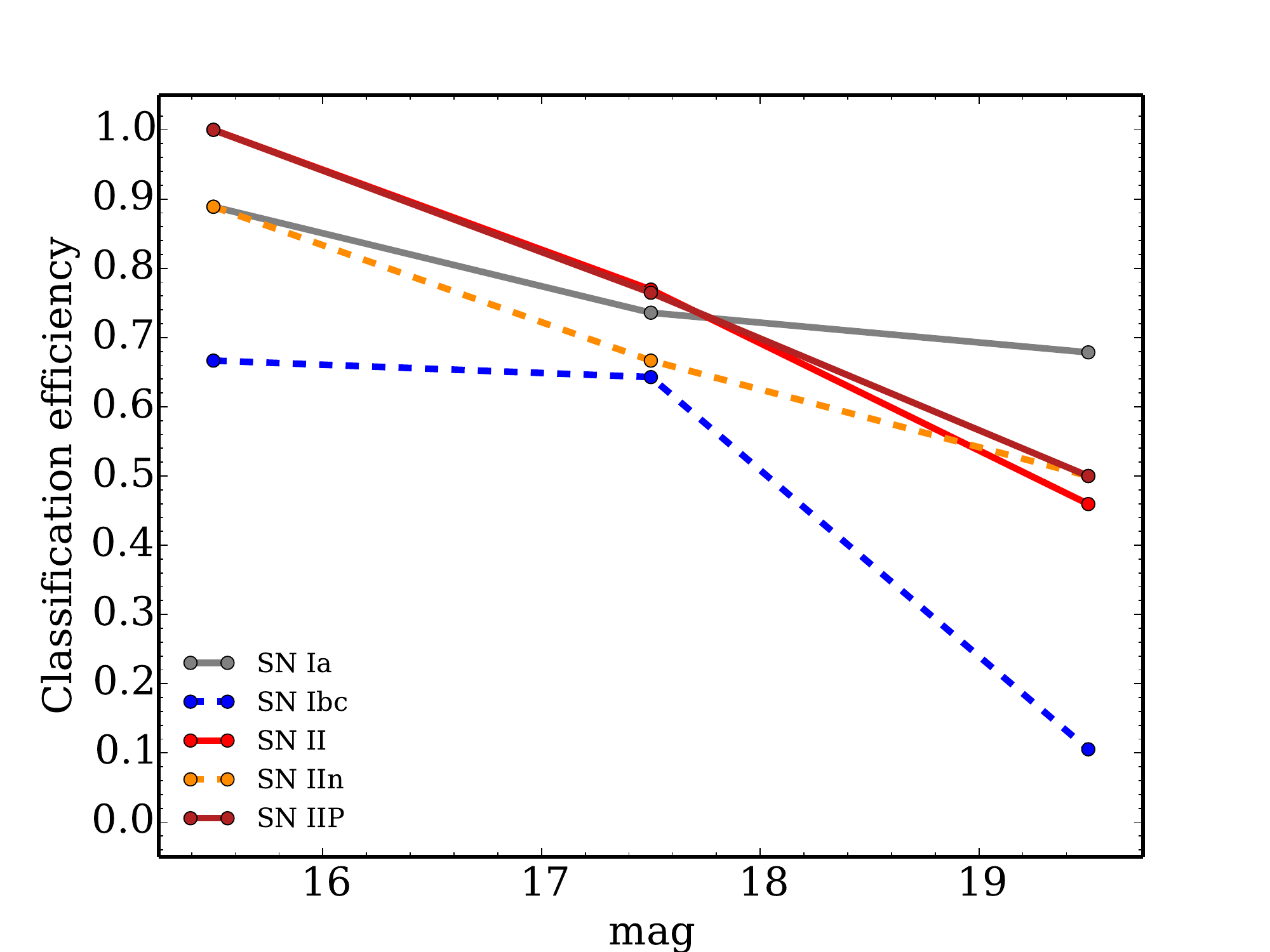} 
\quad
\subfigure{\includegraphics[width=1.1\columnwidth]{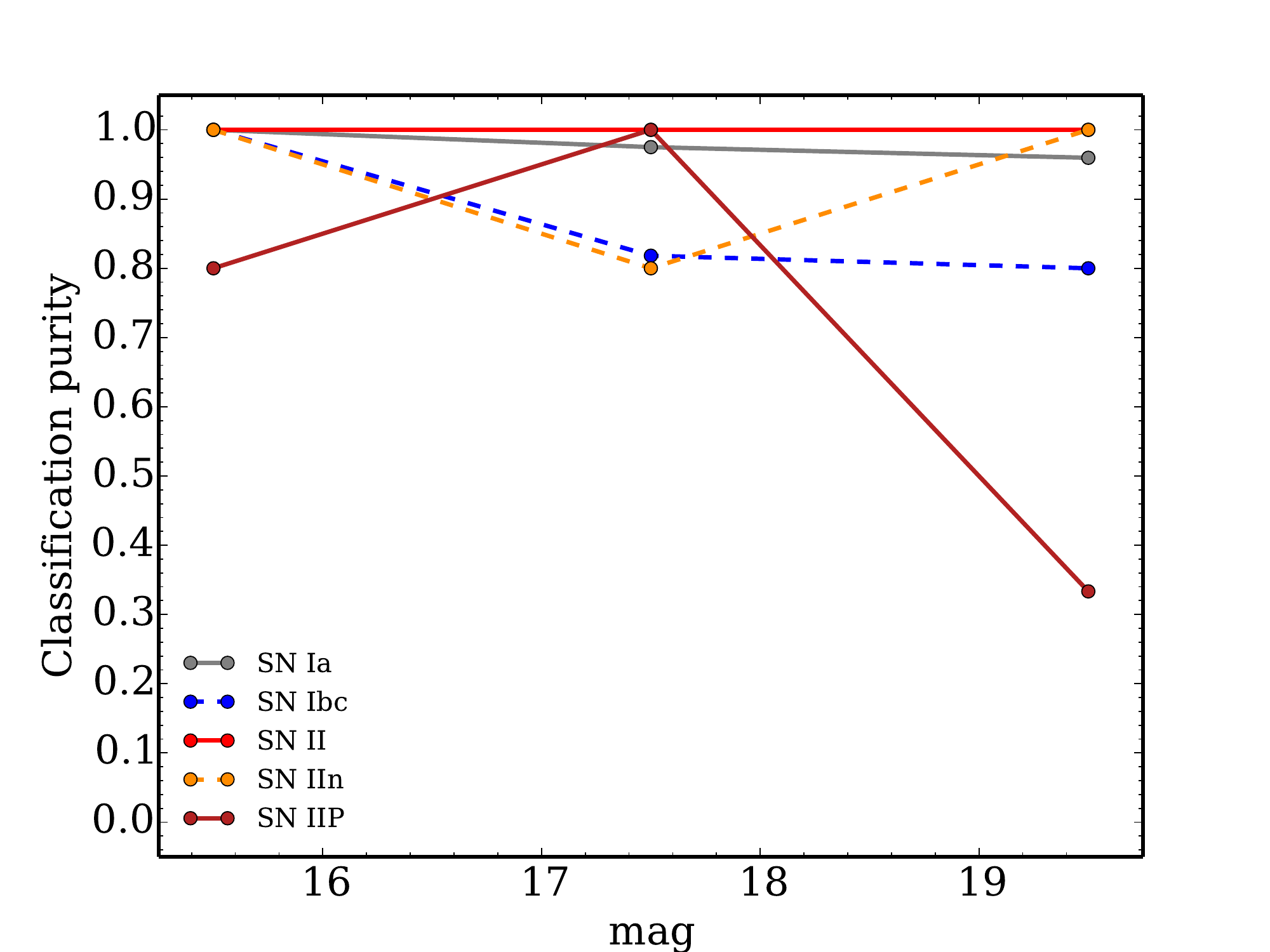} }} }
\mbox{
\subfigure{\includegraphics[width=1.1\columnwidth]{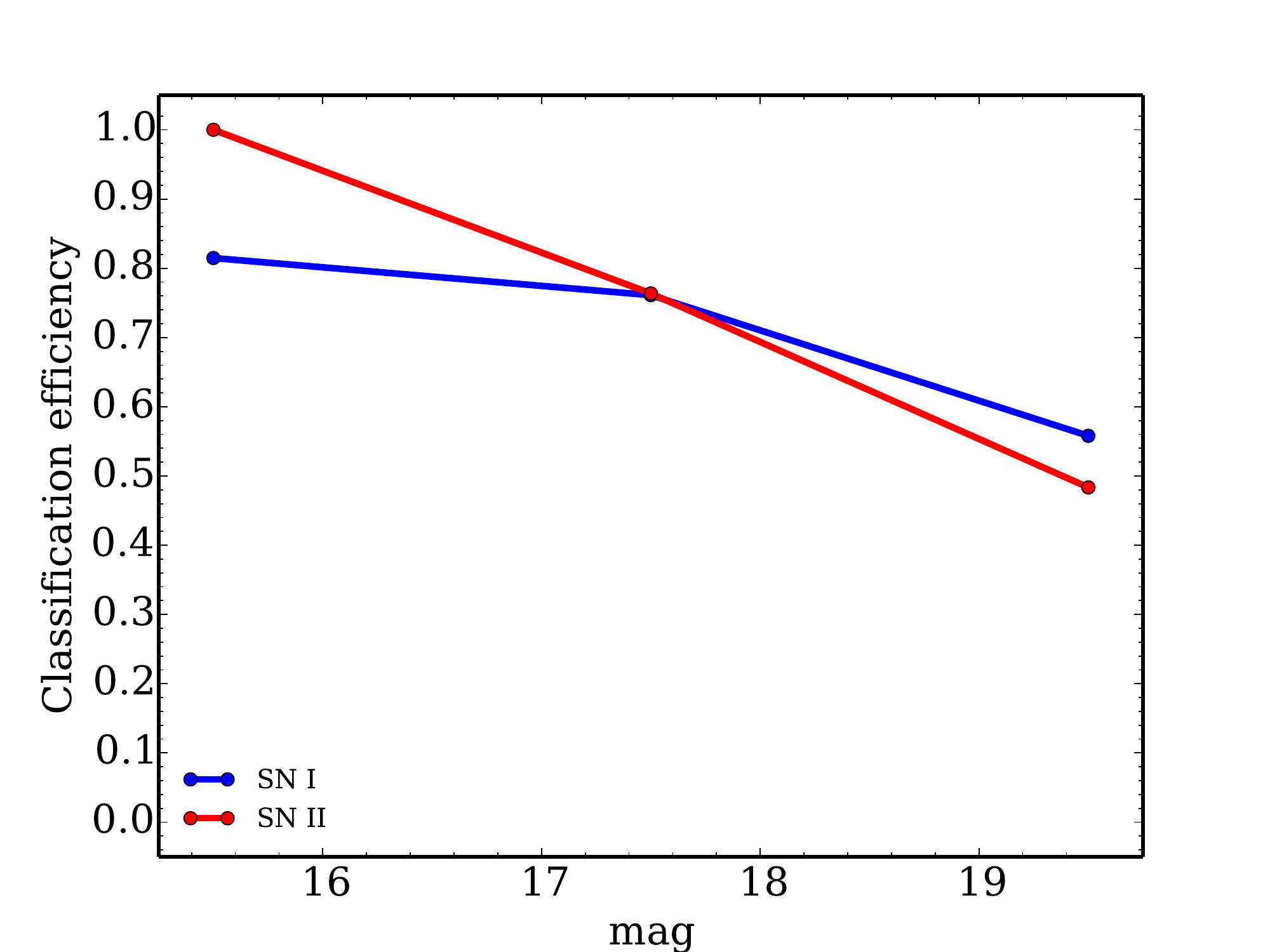}
\quad
\subfigure{\includegraphics[width=1.1\columnwidth]{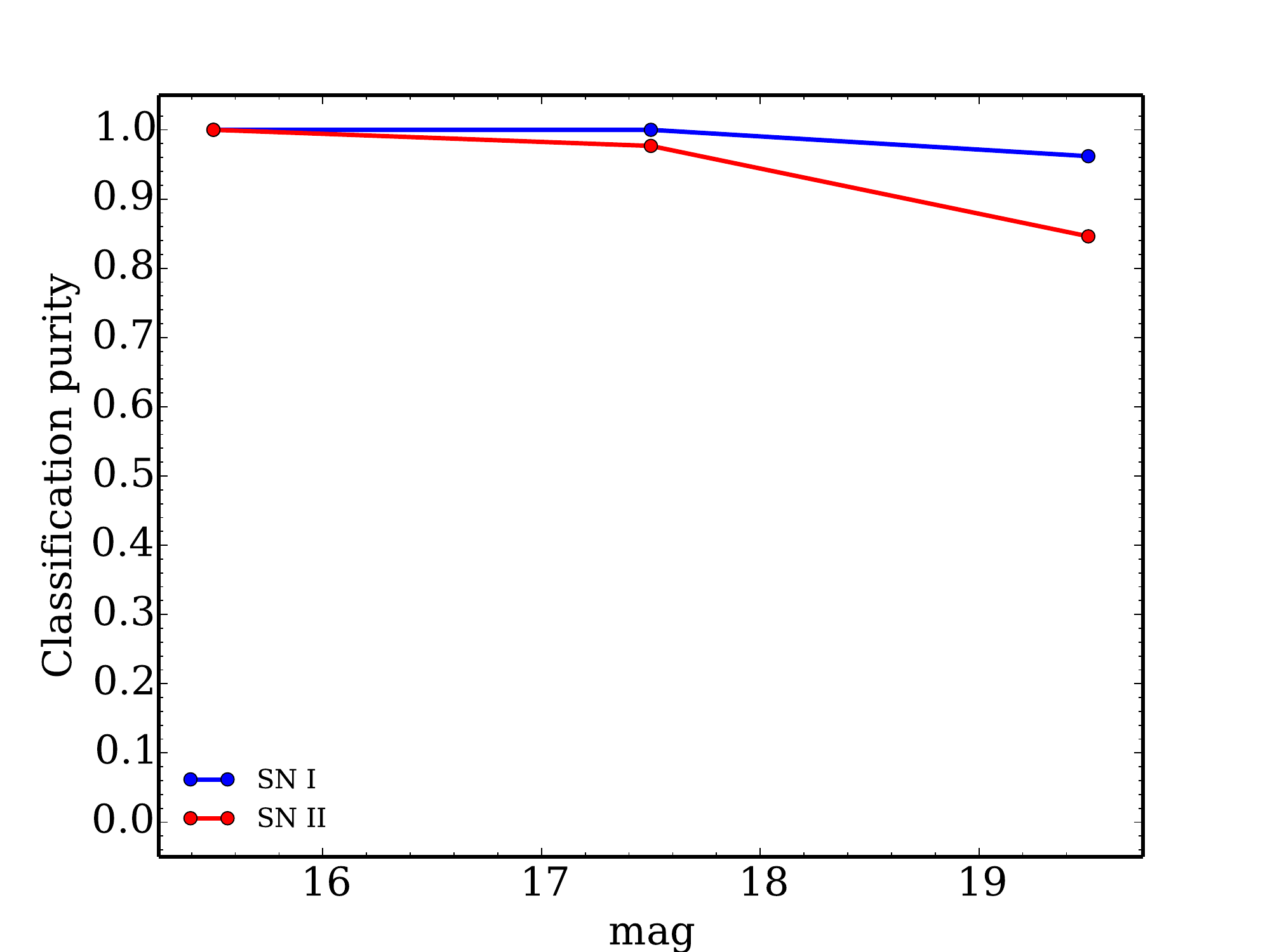} }}
}
\caption{{\it Top left}: Classification completeness (efficiency) as a function of magnitude for each individual class for the PESSTO dataset. {\it Top right}: Classification purity as a function of magnitude and class. 
{\it Bottom left}: Classification completeness as a function of magnitude for the two main SNe subtypes for the PESSTO dataset. {\it Bottom right}: Classification purity as a function of magnitude and class for the two main SNe subtypes for the PESSTO dataset.  The generally disjoint behaviour of the curves is due to the low number of objects per magnitude bin for the PESSTO dataset.
}
\label{fig:accuracy_pessto}
\end{figure*}

We ran the GS-TEC code to classify the PESSTO targets generally obtaining similar results to those from the cross-validation test. In the PESSTO case the accuracy and purity have been extended to magnitude $G$=15 as there were some objects populating that magnitude range. However, due to the low number of objects per class in each magnitude bin, we decided to present the performance binned in intervals of 2 magnitudes, instead of one.

Figure \ref{fig:accuracy_pessto} shows the classification efficiency and purity. This test is more realistic than the ones described previously as the quality of the spectra to be classified is directly associated with brightness and redshift. As PESSTO transients were selected from a realistic survey we did not use the weights in the purity calculation, as the ratio between objects belonging to different classes is already implicit in the test sample.

\begin{figure*}
\centering
  \subfigure{\includegraphics[width=1.6\columnwidth]{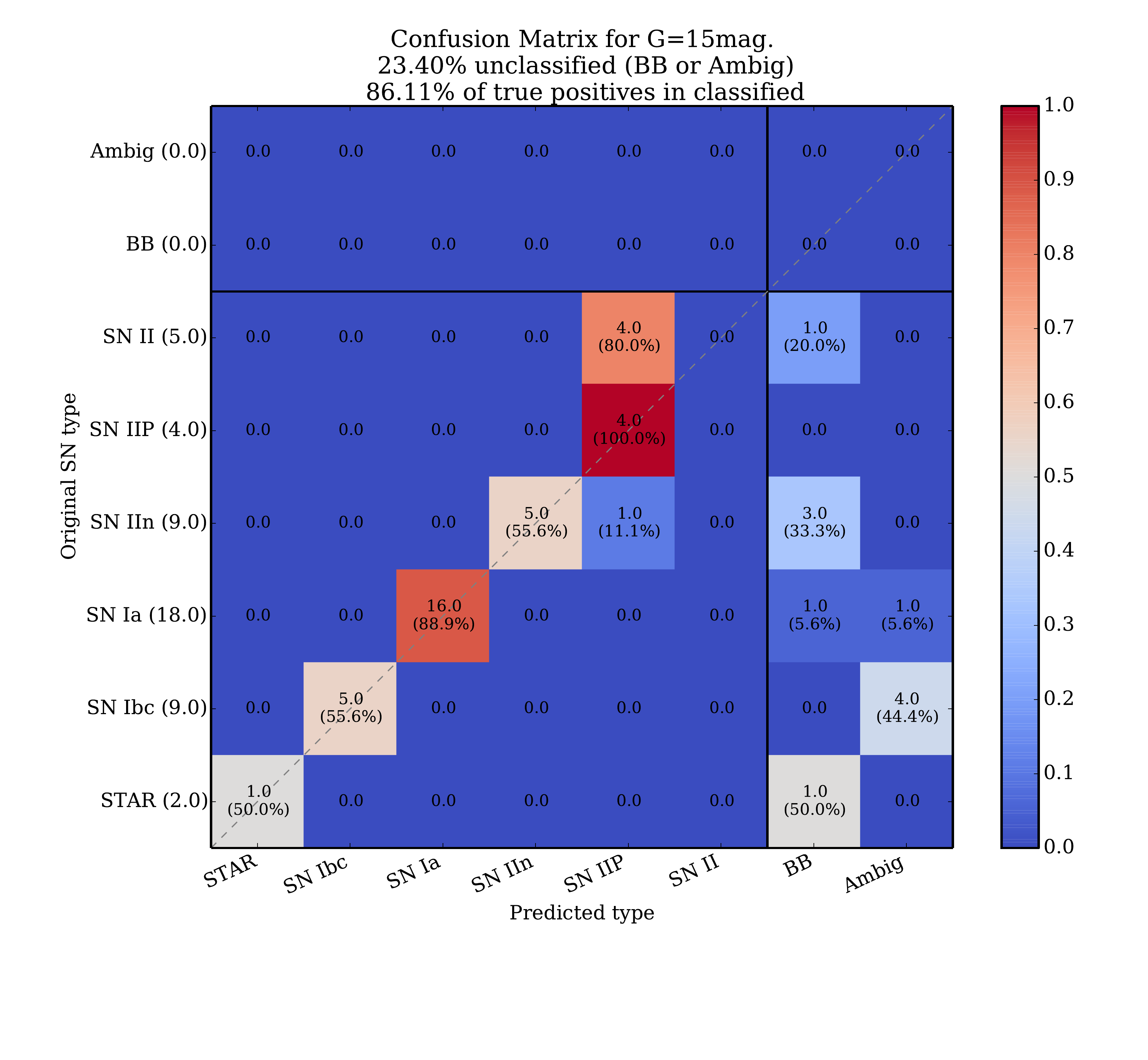}}\\ \vspace{-7.5em}%
  \subfigure{\includegraphics[width=1.6\columnwidth]{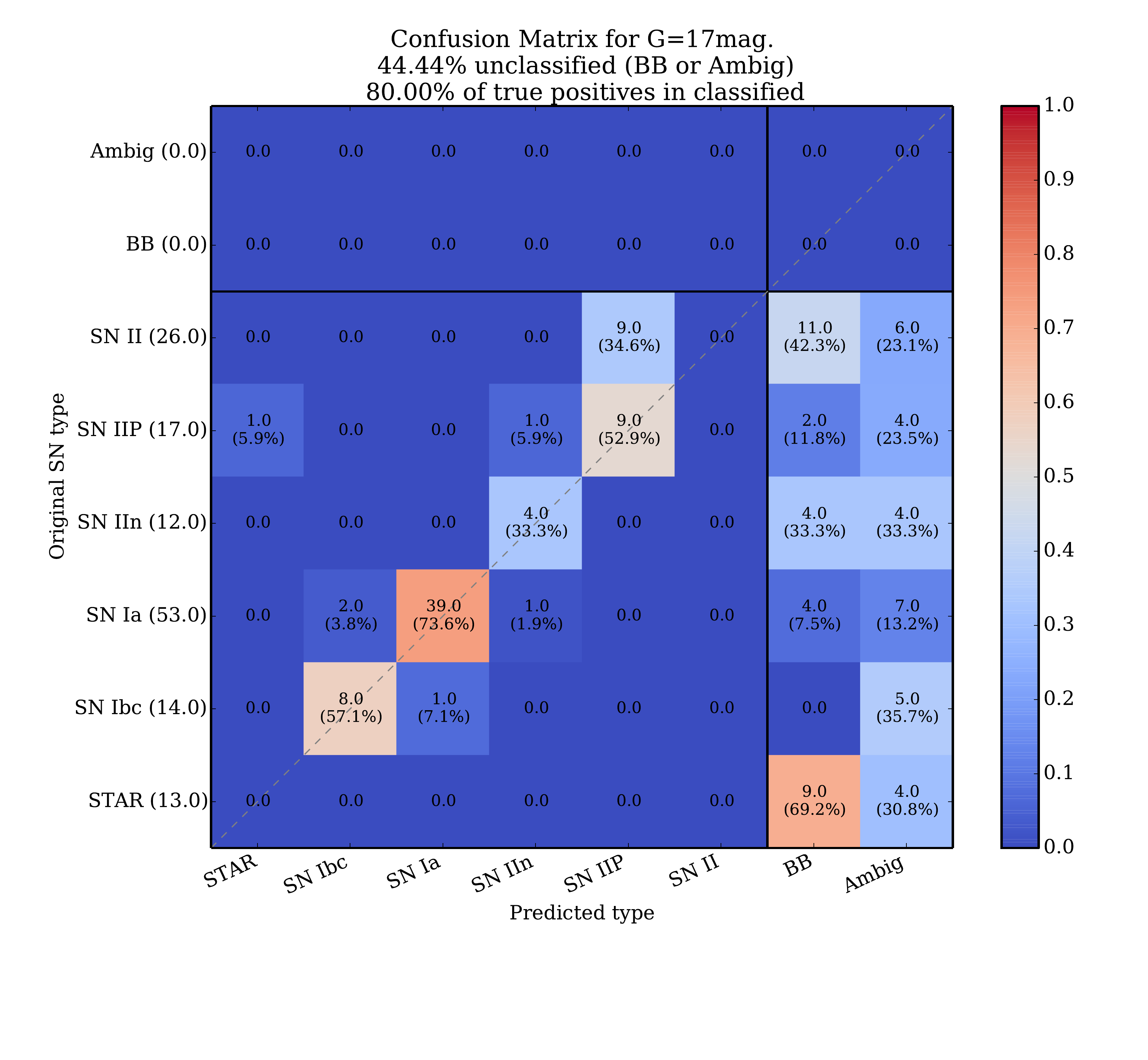}}
  \vspace{-7em}
      \caption{Confusion matrices for the bright end. The X axis represents the class type predicted by the classifier and the Y axis represents the true type. The number in parenthesis indicates the number of spectra used in the test set. The percentages are given relative to this number. The black line separates the real types from the artificial types: BB and Ambiguous. The color bar indicates the percentage of objects that belong to each category.}
 \label{fig:confumat_bright_pessto}
\end{figure*}

\begin{figure*}
\centering
  \subfigure{\includegraphics[width=1.6\columnwidth]{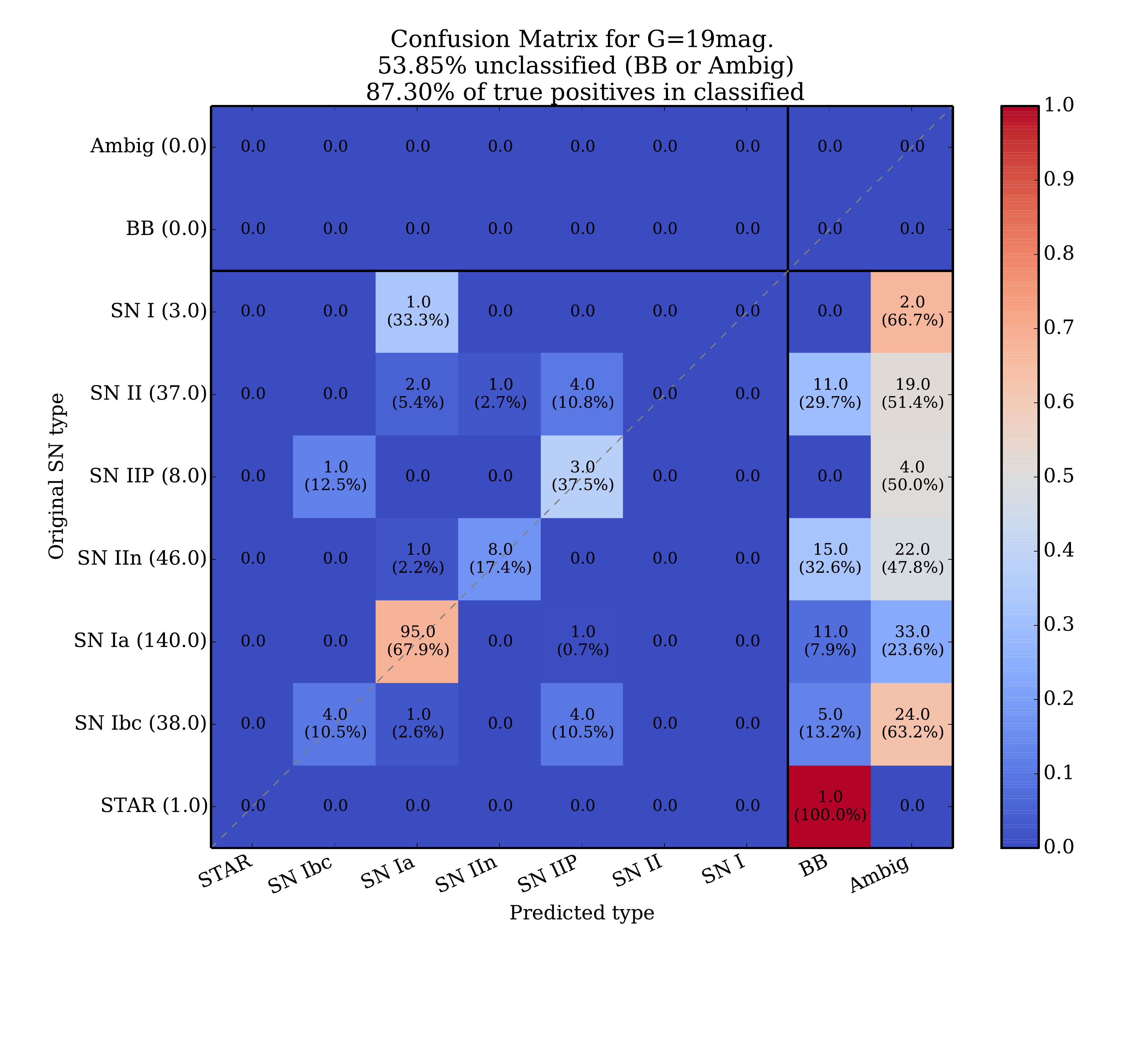}}
    \caption{Confusion matrices for the faint end. The X axis represents the class type predicted by the classifier and the Y axis represents the true type. The number in parenthesis indicates the number of spectra used in the test set. The percentages are relative to this number. The black line separates the real types from the artificial types: BB and Ambiguous. The color bar indicates the percentage of objects that belong to each category.}
        \label{fig:confumat_faint_pessto}
\end{figure*}

On average we see that GS-TEC performs well in recognizing the standard SNe types. However, the confusion matrices in Figure \ref{fig:confumat_bright_pessto} and Figure \ref{fig:confumat_faint_pessto} show some misclassified objects as well. Visual inspection of the high resolution spectra for these problematic cases shows that these happen in the case of specific particular types (SNe 91bg for example), narrow emission lines or poor signal-to-noise in the original high-resolution spectra. This demonstrate that our system provides a useful tool to recognize the most standard SNe types and to estimate their redshifts.

\begin{figure}
\hspace{-1cm}
\subfigure{\includegraphics[width=1.2\columnwidth]{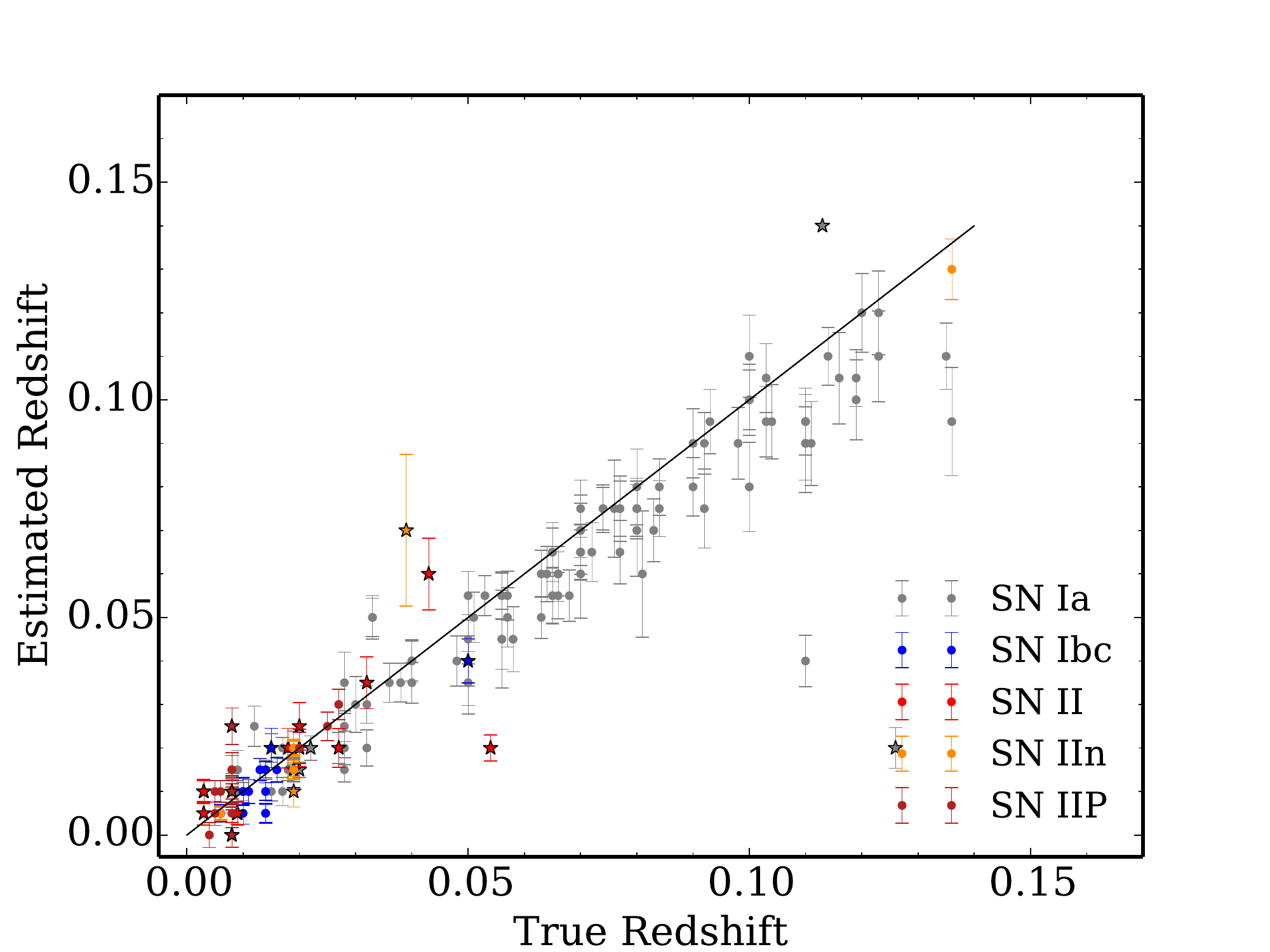} }

\caption{Performance of redshift parameter estimation for the PESSTO dataset. Estimated values for redshift are plotted against the true values from the spectral archive.
}
\label{fig:parameters_pessto}
\end{figure}

Parameter estimation for the PESSTO dataset can only test the redshift estimation as there is no reliable information available on the object epoch. Figure \ref{fig:parameters_pessto} shows the redshift estimation. The tests show that it can be retrieved with an accuracy of $\sigma_z \leq 0.008$ for SNe type II and $\sigma_z \leq 0.013$ for SNe type I. The scatter plot shows that the redshift, specially on the faint end, is slightly biased towards lower values. This can be explained by the fact that the magnitude estimation method used slightly overestimates the magnitude of the transients.


\section{Discussion}\label{sec:discussion}

\subsection{Possible improvements during the mission }

The use of ancillary data from the Gaia Science Alerts process allows the possibility of including additional information on the object, such as the previous classification in the case of longer term variable objects, colors in additional bands and the object environment, for example presence of a nearby galaxy, its type and color. In this context, although GS-TEC can be understood as an independent system it can readily be used to contrast, complement and expand the information provided by parallel modules.

Moreover, the \gaia{} deterministic scanning law makes it easy to check the last date when the satellite was pointing at the transient location giving the last non-detection time.  That information can be used to set an upper limit for the transient epoch and help to restrict the parameter space.

Finally, the most important improvement will come after several months of data compilation in the \gaia{} format, when the newly acquired data, once confirmed by the ground based follow-up resources, will be added to the reference library. This new data will gradually create the ultimate training set for GS-TEC. Having a big training set with real (not simulated) data format is expected to provide highest improvement for the classification performance \citep{Brink2013}.

\subsection{Comparison with other classifiers}

Spectral SNe classification is not a new problem. There are several high resolution SNe SED classifiers, such as SNID: \cite{2011ascl.soft07001B}, GELATO \citep{Gelato2008} or Superfit \citep{Howell2005}. These codes base their classification strategy on comparing the input medium-high resolution spectra with a collection of individual object spectra. The core approach for these codes is to fit and subtract a continuum to remove the possible flux calibration and reddening effects and compare the remaining lines. In order to use these codes the spectra need to have enough signal-to-noise and resolution to distinguish the main spectral features. This kind of approach is difficult to apply to a case like \gaia{}, where the spectral resolution is variable, the spectra are segmented into two parts and the median signal-to-noise is around 10 for magnitude 17 and 2 for the fainter 20 mag. It is untenable to use these approaches, or even a similar strategy, for \gaia{} transient classification. Provisional tests indicate that these kind of solutions only work for very good signal-to-noise spectra with magnitudes 
around 16 or 17.

In contrast, GS-TEC has been designed to work within the \gaia{} instrumental reference framework, whereby the continuum shape of the spectrum plays an important role. Our approach also make use of additional information, such as the object magnitude and generic class type characteristics to achieve a more robust solution.  Our main goal is to create an automated discovery and identification process for the most common, and standard, transient types, leaving ground-based follow-up to provide additional information about possibly interesting \textit{ambiguous} or \textit{black-body}-like spectra.

\section{Summary and Conclusions }\label{sec:conclusions}

We have presented an algorithm for processing \textit{Gaia} low-resolution spectrophotometric data that is capable of estimating the main class of a transient event and some of its non-intrinsic parameters, such as the redshift and epoch of the explosion. The algorithm has been tested on a set of ground-based observations which presented high heterogeneity among types and epochs.

The conclusion from the current work are summarized as follows:
\begin{itemize}
\item \textit{Gaia} low-resolution spectrophotometric and broadband photometric data, coupled with realistic priors, carries enough information to be used for classification of transients;
\item GS-TEC has proven to be an efficient independent module to obtain accurate information on transient class and parameters particularly for SNe having standard spectral shapes and strong features;
\item the efficiency of classification is 85\% at the bright end for SNe type I and 76\% for SNe type II. However, it decreases to 60\% and 48\% respectively for magnitude 19. Class purity is 98\% and 90\% at the bright end for SNe type I and SNe type II, then it decreases to 95\% and 84\% for objects at magnitude 19;
\item redshifts for both main types of SNe can be predicted with an accuracy $\sigma_z \lesssim 0.01$;
\item the main source of confusion at bright magnitudes are variable stars. 
However, this should not be a major problem since nearby SNe are a minority, and they will be promptly discovered and characterized by ground-based observing facilities;
\item for fainter magnitudes the highest confusion comes from within similar SNe types, the group SN I and SN II, as they have similar spectral features and which cause confusion at low signal-to-noise. Providing a more general classification type increases our confidence in the result.
\end{itemize}

Ground-based surveys that collaborate with \textit{Gaia} will benefit
from our module as it will provide additional information on the transient object nature, which may enable more efficient filtering of alerts and therefore better resource allocation for follow-up.

\section{Acknowledgements}
The research leading to these results has received funding from the European Union Seventh Framework Programme ([FP7/2007-2013] under grant agreement num. 264895. 

This research has made use of the CfA Supernova Archive, which is funded in
part by the National Science Foundation through grant AST 0907903.

Based on observations collected at the European Organisation for Astronomical Research in the Southern Hemisphere, Chile as part of PESSTO, (the Public ESO Spectroscopic Survey for Transient Objects Survey) ESO program ID 188.D-3003.

Thanks to Vasily Belokurov for useful discussion and comments, to Ofer Yaron, for his help with the WiseRep repository data and the Padova-Asiago SN group for providing data and useful comments.

\label{lastpage}

\bibliographystyle{mn2e}
\bibliography{mnras}

\bsp

\end{document}